\documentclass[letterpaper,twocolumn,unpublished,11pt]{quantumarticle}
\pdfoutput=1
\usepackage[english]{babel}

\usepackage[T1]{fontenc}
\usepackage[utf8]{inputenc}
\IfFileExists{microtype.sty}{\usepackage{microtype}}{}

\usepackage[allcolors=quantumviolet,unicode=true,colorlinks,pdfauthor={Brandon Rodenburg}]{hyperref}
\usepackage{graphicx}
\usepackage{lipsum}

\usepackage{amssymb,amsmath,braket,cancel}
\usepackage{siunitx}
\usepackage{commath} % includes \dif \abs etc

\DeclareMathOperator*{\argmax}{arg\,max} 
 
\usepackage[square, numbers, sort&compress]{natbib}

\usepackage{ifpdf}

\usepackage{multirow}

\usepackage{framed}

\usepackage{qcircuit}

\title{Estimating the Power of a Quantum Computer.}
\author{Brandon Rodenburg}
\affiliation{Quantum Technologies Group, MITRE, 200 Forrestal Rd. Princeton, New Jersey 08540, USA}
\email{brodenburg@mitre.org}
\orcid{0000-0003-2150-0576}
\thanks{\smallskip\newline Approved for Public Release; Distribution Unlimited.
Public Release Case Number 24-3915.
\textcopyright 2024 The MITRE Corporation. ALL RIGHTS RIGHTS RESERVED.} 

\begin{document}
\maketitle
\renewcommand{\sectionautorefname}{Section}

%% Paper sections
% flatex input: [abstract.tex]
\begin{abstract}
Various benchmarking metrics have been developed to quantify the performance of
quantum computing hardware and help evaluate development. However, it is not
always necessary to know the metric values precisely. This is especially true
for potential end-users who may not be experts in the underlying technology
itself. In this work, we show how to estimate the quantum volumetric metrics
defined in~\cite{DHS_paper1} based on system parameters such as qubit number,
qubit layout/connectivity, and physical error rates. As part of this work, we
also include an initial analysis of how the overhead required for quantum error
correction in systems below the error correction viability threshold affects
the metric value of that system.

\end{abstract}

% flatex input end: [abstract.tex]

%% Paper sections
% flatex input: [intro.tex]
% \section{Introduction} %Intro does not need to be in a section
Quantum computing is a nascent technology, but developing rapidly. The center
of mass of this field has shifted from being primarily basic research within
academic and national labs, to large scale commercial efforts largely funded by
venture capitalists and private industry. In the last few years alone, hundreds
of quantum computing startups have appeared in this space. This includes a
significant rise in companies focused on software and applications, rather than
hardware development alone~\cite{QCCommercial2020}. In addition, quantum
computing has caught the eye of big business with 74\% of large global
enterprises having begun adopting quantum computing, with the majority (71\%)
of these companies having quantum computing budgets exceeding one million US
dollars per year~\cite{ZapataEnterpriseAdoption2022}.

The community of those interested in and/or working in the field of quantum
computing has clearly grown beyond simply being the domain of researchers who
are deep subject matter experts in the technology. With this growth comes the
requirement of clear performance metrics that can be used and understood by
both experts and end-users alike. Such metrics allow organizations to better
plan for disruption, engage with the technology, and to more clearly discern
hype from opportunity. Volumetric metrics have arisen as a valuable framework
for meeting these needs~\cite{DHS_paper1}, and is the basis for the rest of
this work.

A detailed introduction of quantum volumetric metrics is given
in~\autoref{sec:QVintro}. In~\autoref{sec:EffectiveErrorRate}, we introduce the
concept of an effective error rate and relate this to the volumetric metrics.
How to account for the physical or topological layout and connectivity of
qubits in actual devices we address in~\autoref{sec:connectivity}. The impact
of available gate set and physical error rates on the effective error and thus
metrics are detailed in~\autoref{sec:PhysError}. Finally, \autoref{sec:QEC}
gives an overview of how the tradeoffs inherent in the use of quantum
error correction, including both encoding overheads and magic state
distillation, can be tuned in order to optimize system performance as measured
by our metrics.

% flatex input end: [intro.tex]

%% Paper sections
% flatex input: [QVintro.tex]
\section{Quantum Computing Metrics}\label{sec:QVintro}
Quantum computing metrics represent quantitative figures of merit that are
meant to reflect the capabilities of some given platform. An ideal metric has a
number of ideal features such as:~\cite{DHS_paper1}
\begin{enumerate}
    \item Universal and platform independent
    \item Applicable to both near (NISQ) and long term (FT) systems
    \item Simple enough to be useful and understandable to non-experts
    \item Representative of the computational power needed to execute quantum
        algorithms
\end{enumerate}

One of the most common high-level metrics used today are based upon the quantum
volume metric originally introduced by IBM in 2017~\cite{QV2018, QV2019}. The
quantum volume is meant to capture the ability to randomly sample from the full
Hilbert space of a set of qubits. To be able to randomly access any part of the
$2^n$-dimensional Hilbert space of $n$ qubits, one needs to be able to run a
circuit to a gate depth of order $d\sim\mathcal O(n)$~\cite{FastScramble2008}.
Because of this, the quantum volume was defined as the largest ``square
circuit'' that a device can perform defined in terms of the qubit number $n$
and logical gate depth $d$ as
\begin{equation}
    \text{QV-1} = \argmax_{n\le n_\text{max}}\left(\min\left[n,d(n)\right]\right),
    \label{eqn:QV}
\end{equation}
where $n_\text{max}$ is the maximum qubits available on the machine and $d(n)$
is the largest gate depth applied to $n$ qubits that can be performed
successfully with high probability.\footnote{The quantum volume was initially
defined as $V_Q = \argmax_n\left(\min\left[n,d\right]^2\right),$ representing the
space-time volume of a square $n\times n$ circuit~\cite{QV2018}. This
definition was later changed to the exponential definition $\log_2(V_Q) =
\argmax_n\left(\min\left[n,d\right]\right)$ which is currently used
today~\cite{QV2019}. However, this leads to unreasonably large numbers for even
modestly useful circuits leading to the log of this number to be the value
often quoted. For this reason, we instead adopt the convention defined in
Ref.~\cite{DHS_paper1} using the first order volumetric metric QV-1, which is
related to the IBM quantum volume definition by $\text{QV-1} \equiv
\log_2(V_Q)$.}

\begin{figure}[htpb]
    \centering
    \includegraphics[width=\linewidth]{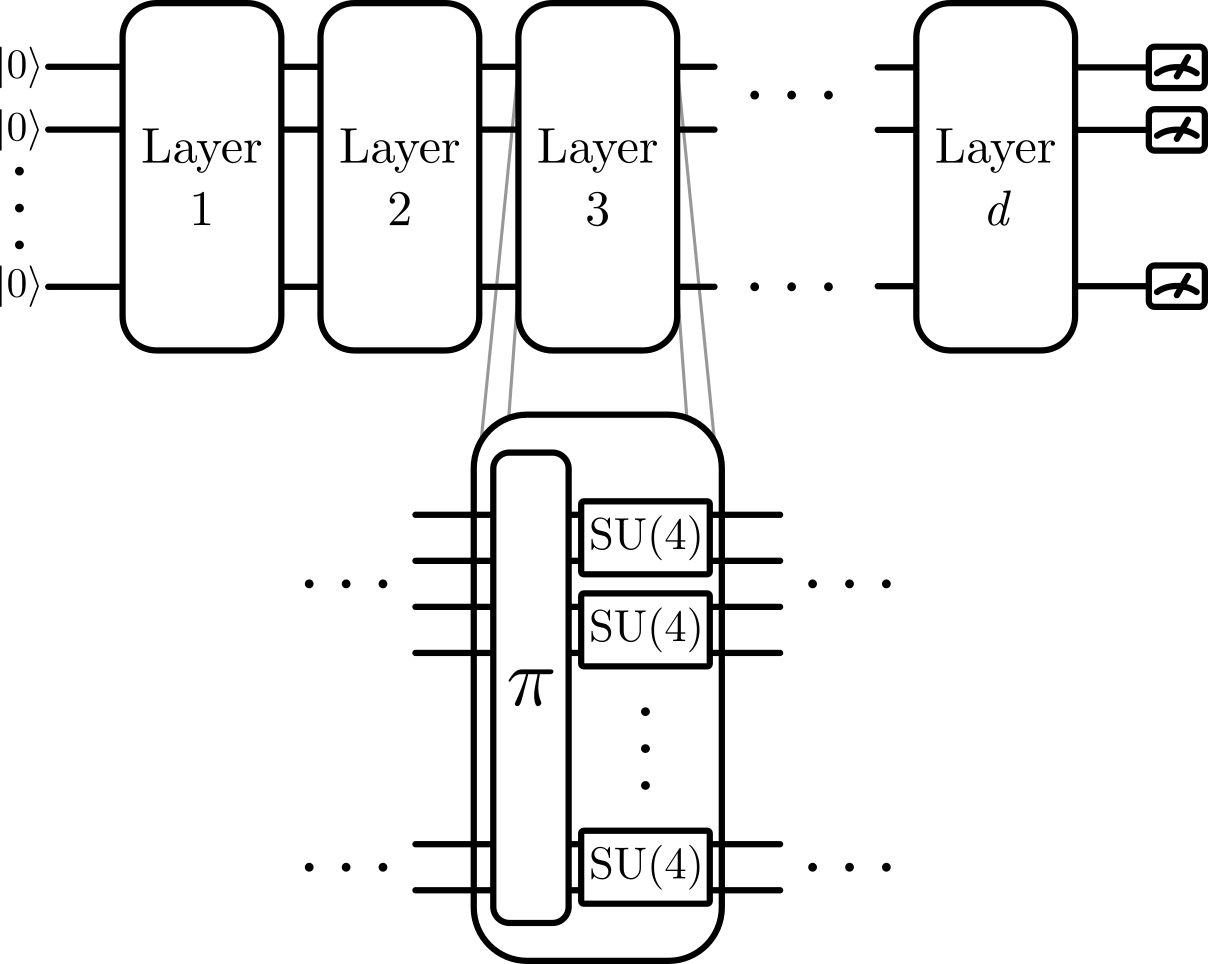}
    \caption{A circuit diagram for benchmarking the quantum volume. The circuit
    consists of $d$ layers of a random permutation of the qubits (represented
    by $\pi$) followed by random two-qubit SU(4) gates.}%
    \label{fig:QVCircuit}
\end{figure}

In order to accurately determine the possible gate depth $d(n)$ for any
subset of $n$ qubits on a quantum computing device, one needs to run a
series of test circuits defined by a benchmark. The original quantum
volume benchmark defines a random test circuit as follows~\cite{QV2018,
QV2019}; each of the $d$ layers of a test circuit consists of a random
permutation of the qubits followed by the pairwise application of random
2-qubit SU(4) matrix as shown in~\autoref{fig:QVCircuit}. The qubit permutation
may consist of sets of swap gates to move qubit states around, or a simple
logical relabeling of the qubits without applying any additional gates to the
extent the qubit layout and connectivity allows.

Despite the elegance of the original quantum volume metric, it is not obviously
connected to specific applications that end-users will care about because most
algorithms do not conform to the square circuit shape. In fact, many
applications require circuit depths that are much deeper than their width or
qubit number~\cite{DHS_paper1}. A solution for this issue has been to move to a
more general framework that treats qubit number and depth as two distinct
parameters, i.e. in terms of a pass/fail table for a large variety of different
qubit number and gate depth combinations~\cite{QVolumetricMetrics2020}. In this
way, one can more directly relate to the end-user application space as quantum
computers can be benchmarked against the actual resource needs of specific
applications~\cite{MirrorCircuitBenchmarking2021,
QEDC_ApplicationBenchmarks2021, EntanglementBenchmark2022}.

\begin{figure*}[htpb]
    \centering
    \includegraphics[width=\textwidth]{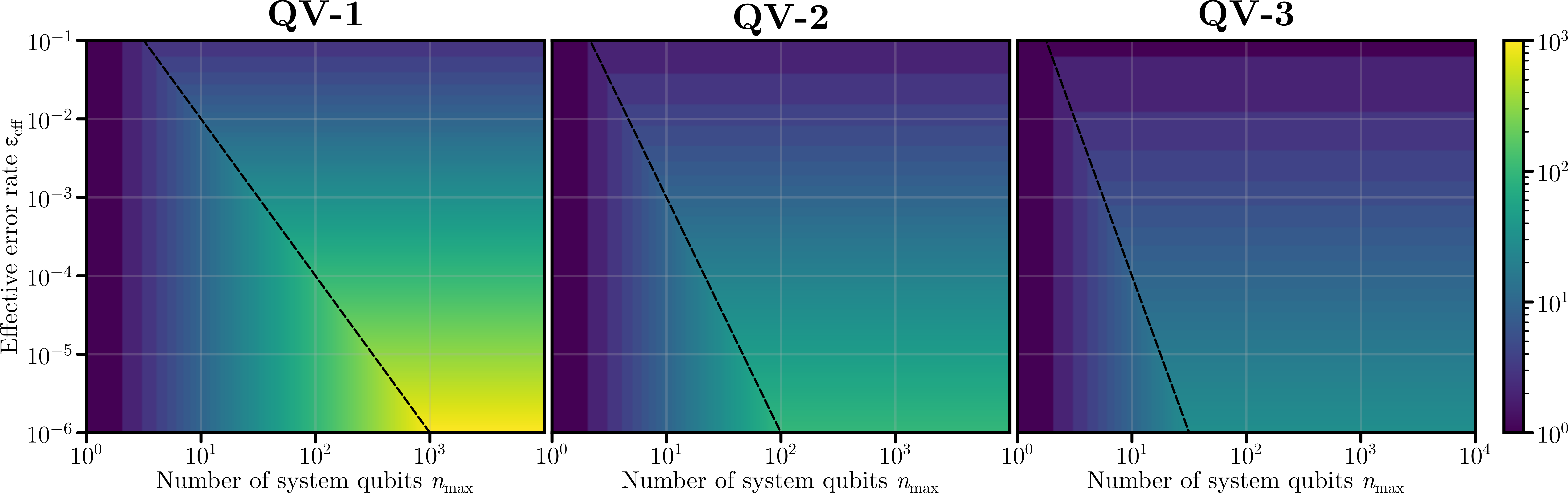}
    \caption{The Quantum Volumetric Class values, QV-$k$, for the first
    three $k$ classes as a function of both system qubit number
    $n_\text{max}$ and effective error rate $\epsilon_\text{eff}$. The
    dashed line in each plot represents the boundary between qubit
    number limited and error rate limited performance.}
    \label{fig:QV_2D}
\end{figure*}

In order to distill the information into a form that is more useful as a
metric, we consider a reduced subset of possible circuit shapes rather than a
table of every possible qubit and depth value. Following the convention defined
in Ref.~\cite{DHS_paper1}, we consider in this work metrics defined by a small
family of circuit shapes, rather than any possible combination of qubit number
and gate depth. Rather than just a square $n\times n$ circuit defined in
\autoref{eqn:QV}, we also consider circuits whose depths scale as $n^2$ or
$n^3$, or in general, $n^k$ which define a family of volumetric metrics.
Mathematically, these additional ``Quantum Volumetric Classes'' (QV Classes)
can be written as
\begin{equation}
\begin{split}
    % \text{QV-1} &= \argmax_{n\le n_\text{max}}\Big(\min\big[n,d      \big]\Big)\\
    \text{QV-2} &= \argmax_{n\le n_\text{max}}\Big(\min\big[n,d^{1/2}\big]\Big),\\
    \text{QV-3} &= \argmax_{n\le n_\text{max}}\Big(\min\big[n,d^{1/3}\big]\Big),\\
    \text{QV-}k &= \argmax_{n\le n_\text{max}}\Big(\min\big[n,d^{1/k}\big]\Big),
\end{split}
\label{eqn:QVClasses}
\end{equation}
where the QV-2 represents the class of applications whose depth scales as
$d(n)\sim\mathcal O(n^2)$, QV-3 as $d\sim\mathcal O(n^3)$, and in general
QV-$k$ as $d\sim\mathcal O(n^k)$. 

The three QV classes QV-1, QV-2, QV-3 can be directly related to specific
applications and together represent the vast majority of known algorithms.
Shor's algorithm for factoring, which represents one of the hardest known
quantum algorithms, scales linearly with the size of the number to be factored
$n$, but the depth scales like $\mathcal{O}(n^3)$~\cite{ShorAlg1994,
QVolumetricMetrics2020}. Therefore, Shor's algorithm is an example of an
algorithm that falls with the QV-3 class, i.e. the QV-3 value of a device gives
you information about how large of a number one can expect to factor. As
examples of the sort of problems one may find in the other QV classes, various
many-body physics and computational chemistry problems fall within QV-2, while
many machine learning and optimization algorithms are within
QV-1~\cite{DHS_paper1}.

% flatex input end: [QVintro.tex]

%% Paper sections
% flatex input: [errorrates.tex]
\section{Effective error rate}\label{sec:EffectiveErrorRate}
In order to estimate the value of any QV Class QV-$k$ (\autoref{eqn:QVClasses})
for a given system, we first need to be able to estimate the scaling behavior
of the depth $d(n)$. Following Ref.~\cite{QV2018}, we imagine a depth-one
circuit using only one and two-qubit gates applied pairwise to $n$ qubits. The
single step error rate is approximately
\begin{equation}
    \epsilon_\text{1 step} \sim n\epsilon_\text{eff},
    \label{eqn:err_1step}
\end{equation}
where we have defined an effective error rate $\epsilon_\text{eff}$ that
represents the average or typical error rate of applying a random SU(4)
between any two qubits. 

From \autoref{eqn:err_1step} we can estimate the circuit depth for which, on
average, a single error occurs as $d\sim 1/(n\epsilon_\text{eff})$.
Substituting into our expression for the QV Class (\autoref{eqn:QVClasses})
gives
\begin{equation}
\begin{split}
\text{QV-}k 
    &= \argmax_{n\le n_\text{max}}\left(\min\left[n,(1/(n\epsilon_\text{eff}))^{1/k}\right]\right)\\
    &= \min\left[n_\text{max},n_\text{opt}(\epsilon_\text{eff})\right]\\
    &= \min\left[n_\text{max},\epsilon_\text{eff}^{-1/(k+1)}\right],
\end{split}
\label{eqn:QVvsEeff}
\end{equation}
where we have used the fact that $d(n)$ is a monotonic decreasing function of
$n$, while $n$ is a (trivially) increasing function of itself and thus these two
functions have a singly defined crossing point as $n$ is varied. Therefore,
QV-$k$ will always be either qubit number limited or error rate limited
respectively (represented by $n_\text{opt}$) depending on whether
$n_\text{max}$ is above or below $n_\text{opt}$. 

This is demonstrated in \autoref{fig:QV_2D} where the QV-$k$ is plotted as a
function of both qubit number and effective error rate. Dashed lines have been
added at $n_\text{opt}$, which represents the crossover between the qubit
number limited and the error rate limited performance regimes, which occur at 
\begin{equation}
    n_\text{opt}(\epsilon_\text{eff}) = \epsilon_\text{eff}^{-1/(k+1)}.
\end{equation}
In the region below or to the left of the dashed line where
$n_\text{max}<n_\text{opt}$, the system is qubit number limited. This is
reflected in the figure by the fact that QV-$k$ increases if $n_\text{max}$ is
increased, but remains constant if the error $\epsilon_\text{eff}$ is
decreased. Above and to the right of the dashed line, we are in the opposite
regime in which the system performance is error rate limited and thus changing
$n_\text{max}$ does not change performance but improving $\epsilon_\text{eff}$
does.

% flatex input end: [errorrates.tex]

%% Paper sections
% flatex input: [connectivity.tex]
\section{Qubit connectivity}\label{sec:connectivity}
In general, we cannot always straightforwardly apply two-qubit gates between
any two qubits in the system. This is due to the fact that for any given
device, generally not all qubits are directly connected to all other qubits.
For example, superconducting qubits are typically arranged on a two-dimensional
integrated circuit where two-qubit gates are applied only between connected
qubits, which are generally only some subset of neighboring qubits on the
chip~\cite{QV2018}. Even systems that today are fully connected, such as some
trapped ion systems, will likely not remain fully connected as the technology
scales to larger connected networks of ion
traps~\cite{IonQCNetworkTopology2016}.

\begin{figure}[htpb]
    \centering
    \includegraphics[width=\columnwidth]{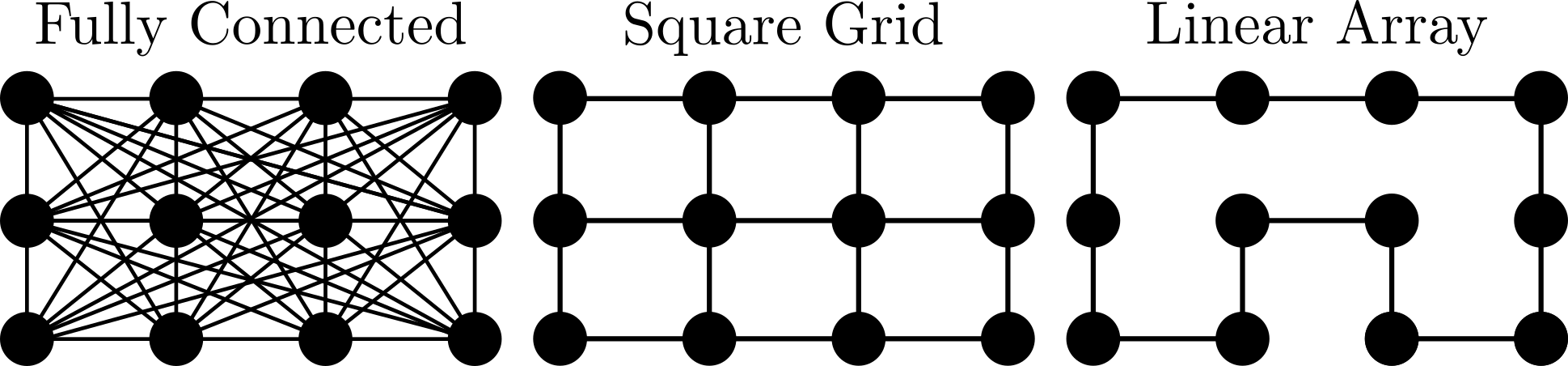}
    \caption{Examples of three different qubit topologies.}
    \label{fig:ConnectivityGraphs}
\end{figure}

Any system whose qubits are not fully connected will require a series of
two-qubit swap gates in order to effectively connect remote qubits. So if we
define the effective two-qubit error rate between connected qubits as
$\epsilon$, but on average need $N$ swaps to connect a random pair of qubits,
then our overall effective error rate is $\epsilon_\text{eff}\sim N\epsilon$.
Now $N$ depends both on the qubit connectivity layout and qubit number $n$. 

\begin{figure}[htpb]
    \centering
    \includegraphics[width=\columnwidth]{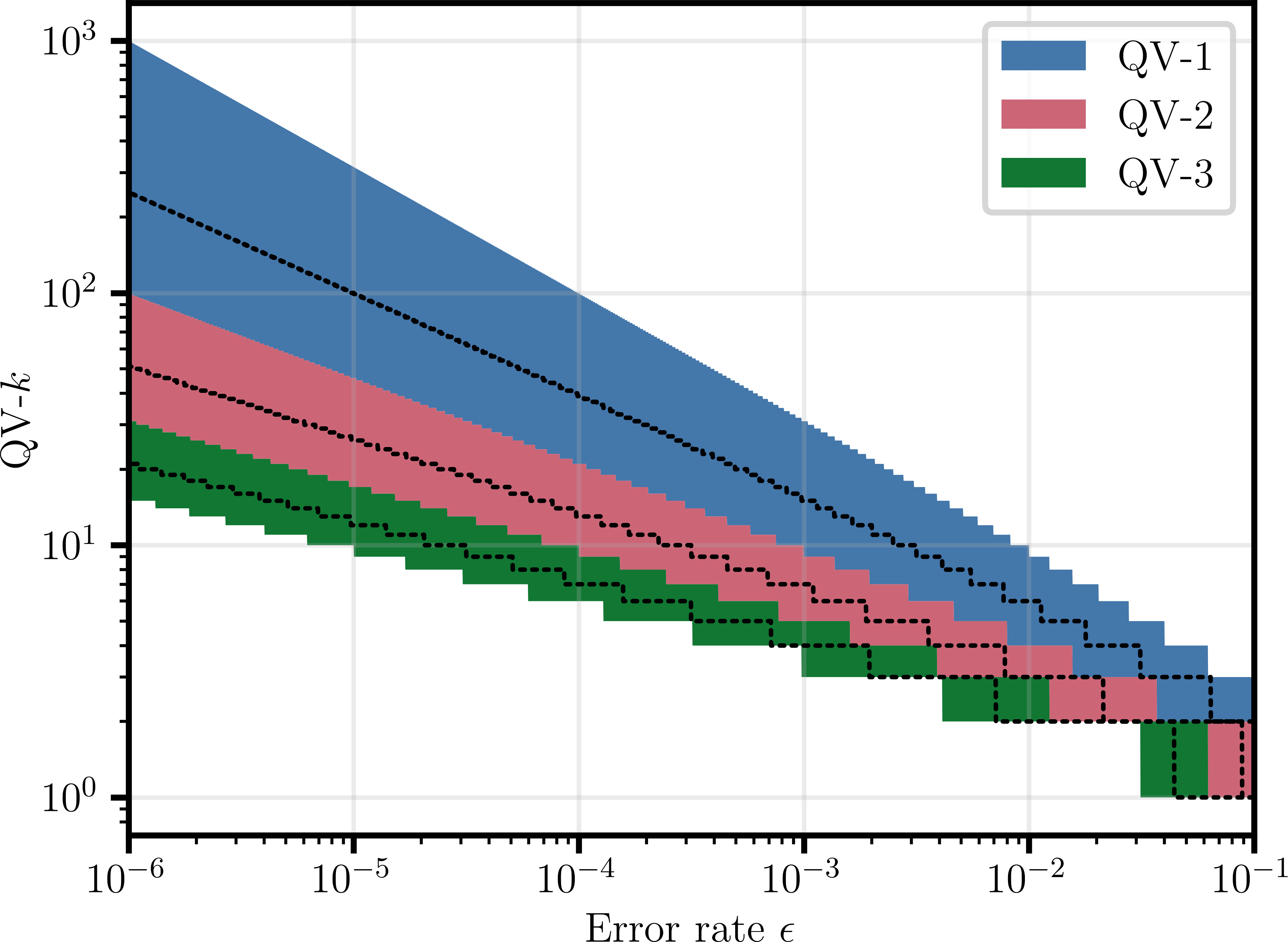}
    \caption{Plot of the three primary QV Classes as a function of the
    connected two-qubit error rate $\epsilon$. Each area represents a range of
    connectivities from $m=0$ (fully connected) to 1 (minimally connected). The
    dashed line within each area represents the intermediate case of a
    nearest-neighbor connected square grid lattice with $m=0.5$.}
    \label{fig:QVk_vs_error}
\end{figure}

Consider the qubit topologies in~\autoref{fig:ConnectivityGraphs}. For a fully
connected graph no swaps are required so $\epsilon_\text{eff} \equiv \epsilon$.
However, a square grid with nearest neighbor connections will require
$N\sim\sqrt{n}$ and a linear array will require $N\sim n$~\cite{QV2018}. We can
parameterize this to any layout by the equation
\begin{equation}
    \epsilon_\text{eff} = n^m\epsilon,
    \label{eqn:eps_eff}
\end{equation}
where the parameter $m$ represents the connectivity. Generally this
parameter will be between 0 and 1, with 0 representing fully connected
systems. We can use this expression to convert QV-$k$ into a function of
$\epsilon$:
\autoref{eqn:QVvsEeff} as a function of 
$\epsilon$
\begin{equation}
\begin{split}
    \text{QV-}k 
        &= \min\left[n_\text{max},n_\text{opt}(\epsilon)\right]\\
        &= \min\left[n_\text{max},\epsilon^{-1/(k+m+1)}\right].
\end{split}
\label{eqn:QVk}
\end{equation}

A plot of the first three QV Classes (\autoref{eqn:QVk}) are shown in
\autoref{fig:QVk_vs_error}. Each QV Class is shown as an area on the graph
representing a range of possible connectivities, with the upper boundary
representing a fully connected system $(m=0)$, and the lower boundary
representing a minimally connected system $(m=1)$. For reference, a dashed line
representing the intermediate case of a connected square grid layout $(m=0.5)$
has been added.

% flatex input end: [connectivity.tex]

%% Paper sections
% flatex input: [physicalerrorrates.tex]
\section{Physical error rates}\label{sec:PhysError}
As detailed in \autoref{sec:connectivity}, the effective error rate,
$\epsilon_\text{eff}$, of applying an arbitrary SU(4) gate between any two
random qubits depends upon qubit layout and connectivity as well as the
effective error rate $\epsilon$ between two directly connected qubits
(\autoref{eqn:eps_eff}). Generally we do not know $\epsilon$ as it is not
directly measured as a physical quantity. However, due to the prevalence of
randomized benchmarking techniques in characterizing devices, we will generally
know the effective error rates associated with the primary physical gates, or
so-called basis gates, that the device can
implement~\cite{RandomizedBenchmarking2008, MirrorCircuitBenchmarking2022}.
This gate set is typically comprised of a set of one qubit gates, together with
an entangling two-qubit gate such as the controlled-NOT or CNOT. Therefore, we
want to find a way of expressing $\epsilon$ (and thus $\epsilon_\text{eff}$) as
a function of these physical error rates. 

\begin{figure}[htpb]
\begin{equation*}
\Qcircuit @C=0.5em @R=1em {
    & \gate{SU(2)} & \targ     & \gate{R_z(\theta_1)} & \ctrl{1} & \qw          & \targ     & \gate{SU(2)}& \qw\\
    & \gate{SU(2)} & \ctrl{-1} & \gate{R_y(\theta_2)} & \targ    & \gate{R_y(\theta_3)} & \ctrl{-1} & \gate{SU(2)}& \qw}
\end{equation*}
    \caption{An arbitrary two-qubit gate is decomposable into at most three
    two-qubit gates and seven single qubit gates.}
    \label{fig:SU4CircuitDiagram}
\end{figure}
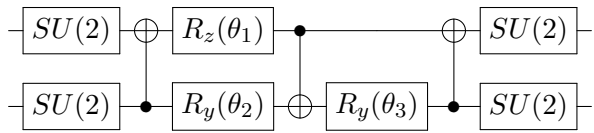

In most modern quantum hardware systems, the two-qubit error rate is
significantly larger than errors from applying single qubit rotations.
Therefore, for simplicity we assume that we can encapsulate all single qubit
gate errors into a single parameter $\epsilon_1$, representing the error rate
associated with applying a general single qubit rotation or SU(2) gate. We then
represent the error rate of the physical two-qubit gate by $\epsilon_2$. We
also assume that errors occur incoherently and are uncorrelated. In principle,
this seems like a limitation to our model. However, in real systems correlated
noise sources are detrimental to the ability to scale quantum hardware and will
thus require techniques to randomize such noise at the hardware
level~\cite{RandomizedCompiling2016, RandomizedCompiling2021}.

\begin{figure}[htpb]
    \centering
    \includegraphics[width=\columnwidth]{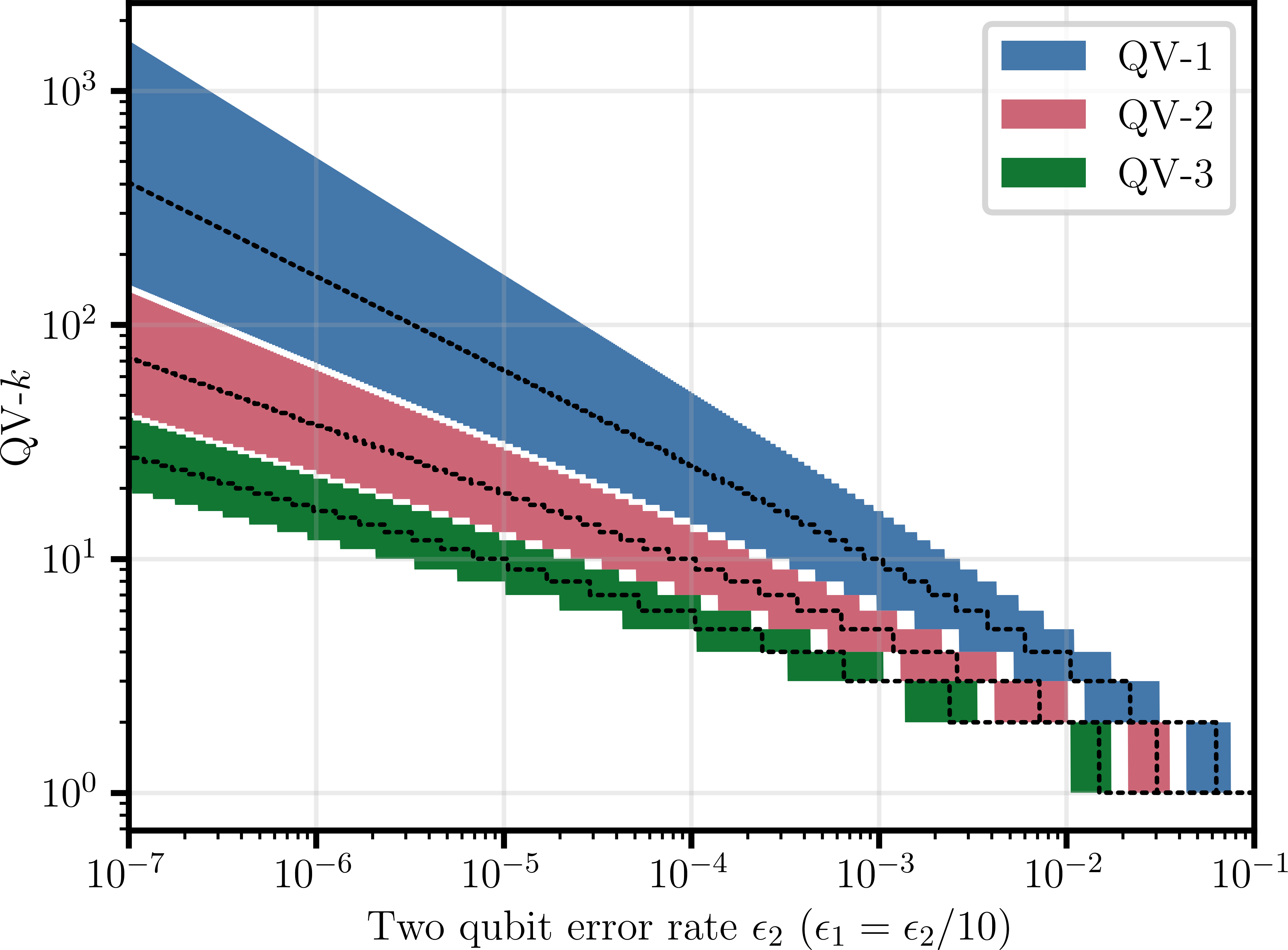}
    \caption{Plot of the three primary QV Classes as a function of the
    physical two-qubit error rate assuming that $\epsilon_2 = 10\epsilon_1$ at
    each point. Each area represents a range of connectivities from $m=0$
    (fully connected) to 1 (minimally connected). The dashed line within each
    area represents the intermediate case of a nearest-neighbor connected
    square grid lattice with $m=0.5$.}
    \label{fig:QVk_vs_physicalerror}
\end{figure}

Any arbitrary two-qubit SU(4) gate can be composed of at most seven single
qubit SU(2) gates (three of which are single axis rotations around $z$ or $y$,
i.e. $R_z$ or $R_y$) and three two-qubit entangling gates, such as CNOTs as
shown in \autoref{fig:SU4CircuitDiagram}~\cite{Optimal2QbGates2004}. Therefore,
the logical error rate $\epsilon$ scales with the physical error rates as
\begin{equation} 
    \epsilon 
    = 1-(1-\epsilon_1)^7(1-\epsilon_2)^3 
    = 7\epsilon_1+3\epsilon_2 +\mathcal O(\epsilon_{1,2}^2),
    \label{eqn:epsilonvsphysicalerror} 
\end{equation} 

In general, \autoref{eqn:epsilonvsphysicalerror} represents an upper bound on
the error rate, as some SU(4) gates will require fewer constituent physical
gates to implement. In particular, implementing a swap gate requires three
CNOTs~\cite{Optimal2QbGates2004}. For devices that are not fully connected, we
expect swap gates to represent a significant portion of the total needed
two-qubit gates. Thus we expect the bound provided by the equation above to be
relatively tight for such systems.

Plots of QV-$k$ are shown in \autoref{fig:QVk_vs_physicalerror} as a function
of $\epsilon_2$. Most current systems are limited by the two qubit error rate
with the single qubit error rates being at least an order of magnitude
smaller~\cite{SupermarQBehcnmark2022}. To represent this, we have assumed that
$\epsilon_1 = \epsilon_2/10$ in the plots.

% flatex input end: [physicalerrorrates.tex]

%% Paper sections
% flatex input: [qec.tex]
\section{Quantum error correction}\label{sec:QEC} 
The analysis in \autoref{sec:PhysError} considered how imperfect gates, if left
uncorrected, limit computational power. Barring revolutionary reductions in
typical gate error rates, quantum computers will need to implement quantum
error correction (QEC) to address realistic problems of interest. QEC is the
ability to actively detect and correct a certain degree of error in the
computation as they occur. However, this ability comes at the cost of
additional resources, in particular, each logical qubit is encoded onto
multiple physical qubits and gates now need to be applied correctly to the
qubit ensemble as a whole. If the error rates including this additional
required overhead are below some threshold, this process results in a logical
error rate lower than the physical one. Although fully fault tolerant operation
below threshold is yet to be fully realized, the field is showing rapid
progress towards this goal~\cite{QEC1, QEC2, QEC3, QEC4, QEC5, QEC6, QEC7,
QEC8, QEC9, QEC10, QEC11}.

In order for our metrics to be maximally useful from an applications
perspective, we follow the convention set in Ref.~\cite{DHS_paper1} and assume
that the QV Metric Classes are defined at the logical level. This means that
when one wishes to determine the metric value QV-$k$ of any system, one can
trade off the effective number of logical qubits for decreased effective error
rate and thus increased maximum circuit depth $d$.

\subsection{Naive QEC Model}\label{subsec:nLvsdc}
If we ignore the overhead needed to perform fault tolerant gates, the QEC gives
us a direct and simple tradeoff. In other words, we replace the physical qubit
number and error rates $n_\text{max}$ and $\epsilon$ in \autoref{eqn:QVvsEeff}
with the logical qubit number $n_L$ and logical error rate $\epsilon_L$ and
optimize the QEC overhead, defined in terms of the number of physical qubits
per logical qubit, to maximize the metric, i.e.
\begin{equation}
    \text{QV-}k 
    = \argmax_{1\le n_L\le n_\text{max}} 
      \left(
          \min\left[n_L,\epsilon_L^{-1/(k+1)}\right] 
      \right).
    \label{eqn:QVk_NQEC}
\end{equation}

We consider the Surface Code to give a specific example of what the $n_L$
versus $\epsilon_L$ tradeoff looks like~\cite{QECSurfaceCode2012}.
The surface code is an example of a stabilizer code defined on a two
dimensional lattice of qubits, which map well onto architectures such as
superconducting systems. In addition, surface codes have relatively high error
threshold rates, on the order of $\sim 1\%$, making these codes very
attractive.

\begin{figure}[htpb]
    \centering
    \includegraphics[width=\columnwidth]{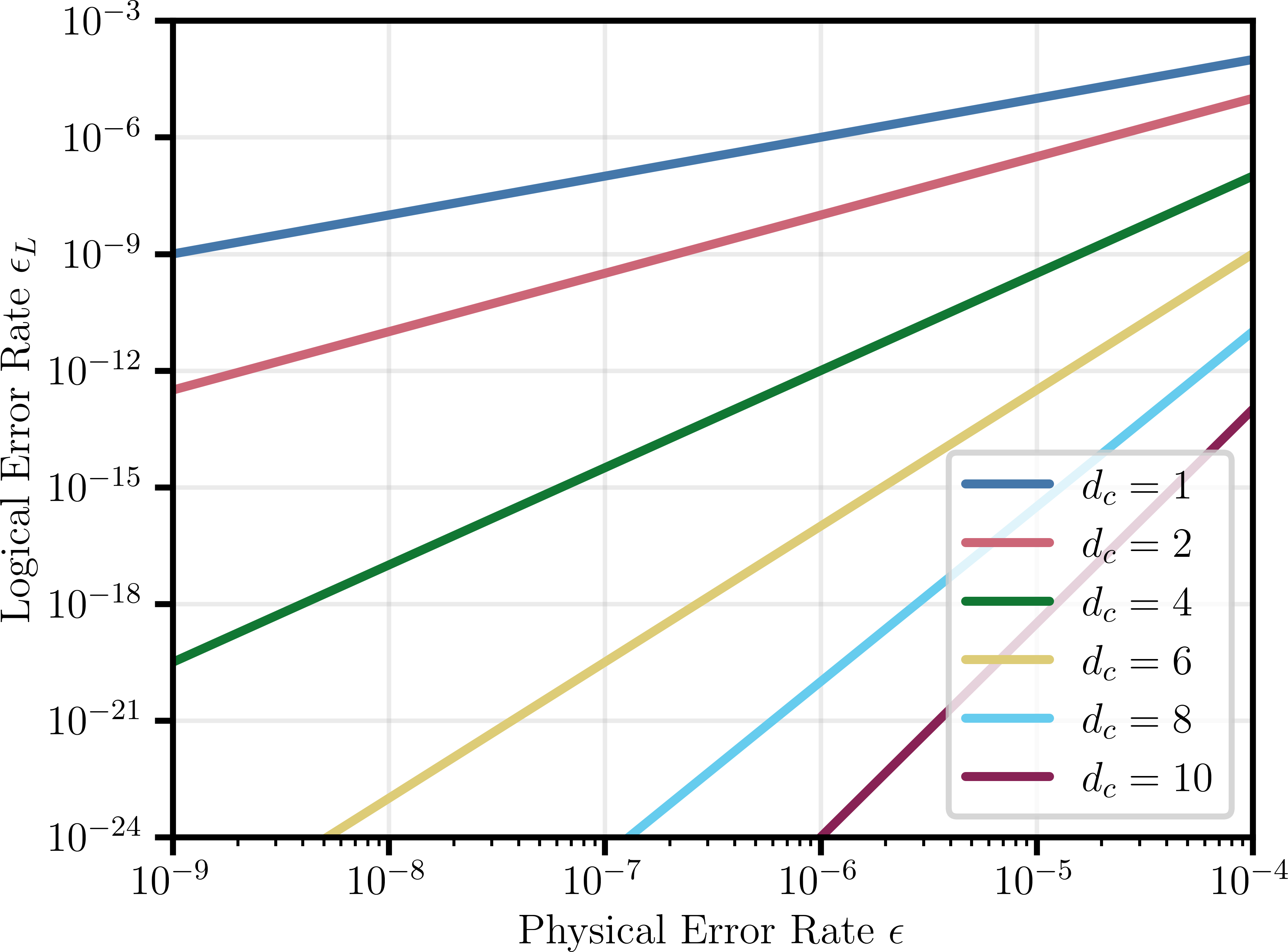}
    \caption{A plot of the logical error rate $\epsilon_L$ for QEC surface
    codes as a function of the physical error rate $\epsilon$ for various
    levels of code distance $d_c$.}
    \label{fig:SC_PhysVsLogicalError}
\end{figure}

\begin{figure*}[htpb]
    \centering
    \includegraphics[width=\textwidth]{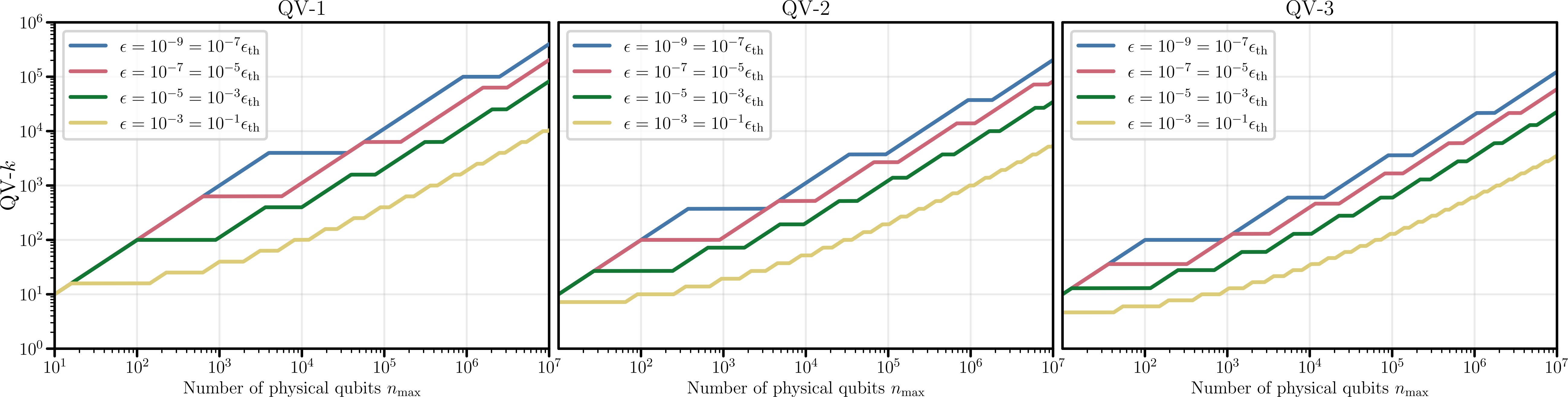}
    \caption{Optimization of the QEC surface code overhead to maximize the
    QV-$k$ for various values of the underlying physical error rate
    $\epsilon$.}
    \label{fig:SC_QVk}
\end{figure*}

QEC codes can be characterized in terms of an integer code distance $d_c$ which
is defined as the number of distinct states or `codes' that exist between
logical bit values. This value is directly related to how many independent
errors can be tolerated before introducing  a logical error. For a given value
of this code distance $d_c = 2t+1$, the QEC code can correct at most $t$ total
errors, i.e., the number of errors must be below $t \le
\lfloor(d_c-1)/2\rfloor$. For surface codes, the number of physical qubits
$n_p$ per logical qubit $n_L$ is given as
\begin{equation}
    \frac{n_p}{n_L} = (2d_c-1)^2.
    \label{eqn:n_LSC}
\end{equation}
We note for the special value of $d_c = 1$, this value is equal to one, i.e.
this represents the case of no error correction. In addition, the logical error
rate can be taken to scale as~\cite{QECSurfaceCode2012,
ForecastingQCTimelines2020}
\begin{equation}
\epsilon_L 
    = \epsilon_\text{th} \left(\frac{\epsilon}{\epsilon_\text{th}}\right)^{(d_c+1)/2},
    \label{eqn:err_LSC}
\end{equation}
where $\epsilon_\text{th}$ is the QEC threshold value. For the purposes of this
work, we will take the threshold to be $\epsilon_\text{th} = 0.01$ for QEC
surface codes, as is generally assumed~\cite{QECSurfaceCode2012,
ForecastingQCTimelines2020}. Plots of how $\epsilon_L$ scales with the physical
error rate $\epsilon$ for various code distances are shown in
\autoref{fig:SC_PhysVsLogicalError}.

Combined, \autoref{eqn:n_LSC} and \autoref{eqn:err_LSC} give us
a simple model we can use for optimizing QEC to maximize QV-$k$. We can think
of both the logical qubit number $n_L$ and error rate $\epsilon_L$ as being
functions of the code distance $d_c$, which reduces \autoref{eqn:QVk_NQEC} to a
one dimensional optimization of the parameter $d_c$. In addition this function
is bounded as the minimum QEC overhead is zero ($d_c\ge 1$), and the number of
available qubits is bounded ($n_p/n_L\le n_\text{max}$). Therefore,
\begin{equation}
    1\le d_c \le \frac{\sqrt{n_\text{max}}+1}{2}.
\end{equation}

The results of this optimization were performed numerically for a variety of
physical error rates and available physical qubits. These results are plotted
in \autoref{fig:SC_QVk}. Every curve of QV-$k$ as a function of $n_\text{max}$
for different physical error rates looks like a monotonically increasing stair
step function with alternating phases of QV-$k$ increasing or remaining
constant respectively. This can be understood by the fact that each discrete
code distance $d_c$ represents a discrete decrease in the logical error rate at
the cost of a discrete increase in physical qubit overhead, $n_p/n_L$. Portions
in which the curves are increasing represent regimes for which QV-$k$ is qubit
number limited and thus increasing the QEC overhead while reducing the error
rate, would decrease the overall performance as quantified by the metric. By
contrast, the plateaus represent regimes in which the performance is error rate
limited and for which there are not enough available qubits to be able to
increase the performance by increasing the QEC overhead.

\subsection{Effect of Imperfect $T$ Gates}
It is a generic feature of quantum error correction that it is not possible to
implement a universal gate set within the code in a straightforward
manner~\cite{EastinKnillNoGo2009}. For instance, one may be able to implement
Clifford gates in a transversal manner, but not small angle rotations such as
$T$ gates. However, not only are Clifford gates not universal, they are also
known to be classically simulatable and thus can offer no real quantum
advantage~\cite{GottesmanKnillTheorem1998}!

A universal fault tolerant gate set is possible, but requires additional
overhead. There are a variety of methods of creating a universal gate
set~\cite{FTQCReview2017}. One of the more promising methods is the use of
magic states which are then consumed in order to implement a $T$ gates (or
possibly other non-Clifford gates~\cite{GidneyCCZMSD2019}) via gate
teleportation~\cite{MagicStateDistillation2005}. The primary overhead required
in this method is the creation of magic states of sufficient fidelity to be
useful for computation. Fault tolerant magic states are generated by combining
lower fidelity magic states in a process known as magic state
distillation~\cite{GidneyCCZMSD2019, MagicStateDistillation2005,
GameOfSurfaceCodes2019, LitinskiMSD2019}

\begin{figure}[htpb]
    \centering
    \includegraphics[width=\columnwidth]{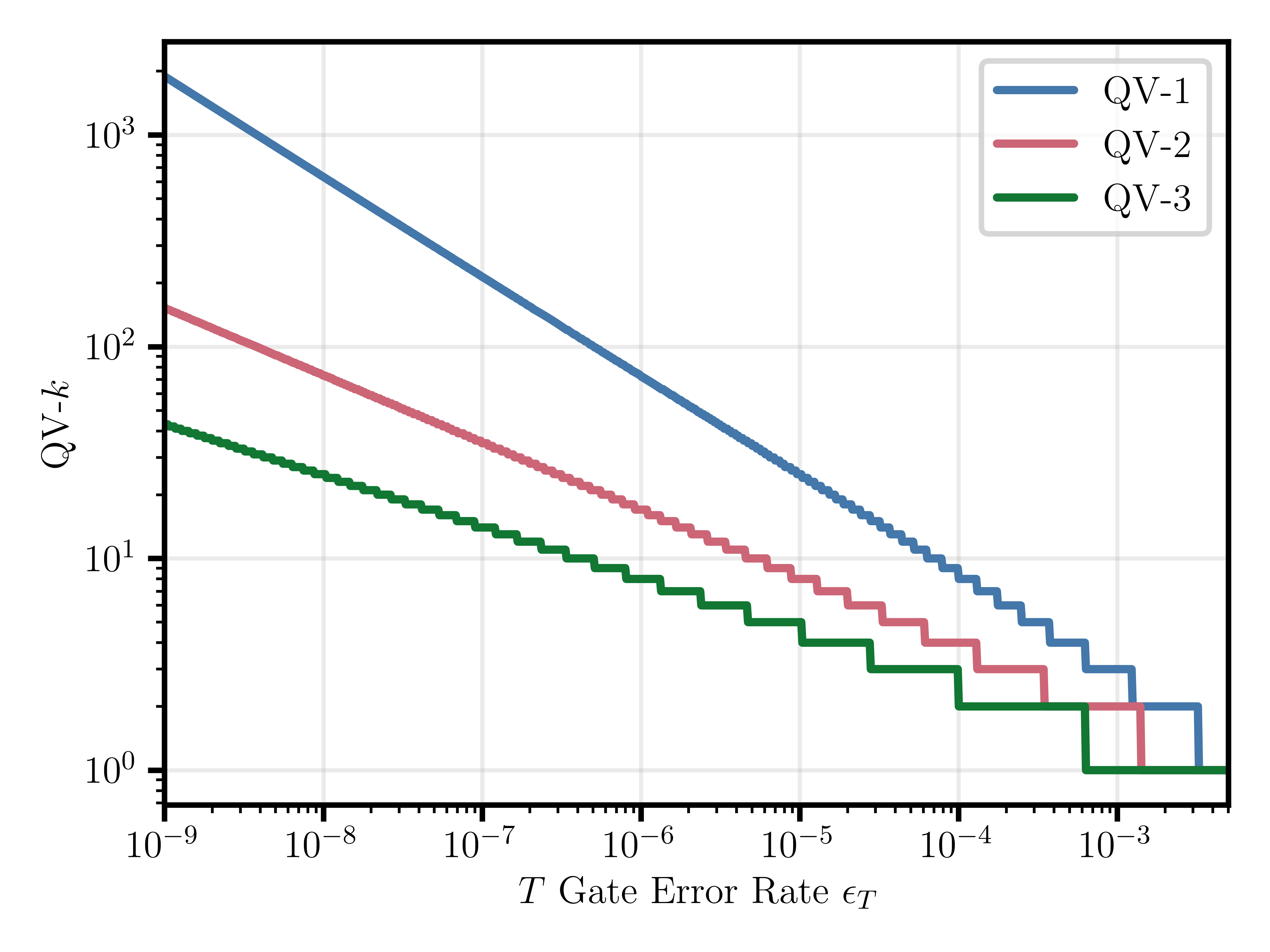}
    \caption{Plot of the three primary QV Classes as a function of the $T$ gate
        error rate $\epsilon_T$. Plot assumes that the error from $T$ gates is
        the dominant error (i.e. $\epsilon_T \gg \epsilon_L$) and that the
        system is error rather limited (rather than qubit limited) for all
        plotted values. The precision $\epsilon_P = 3\epsilon_T/\ln(2)
        \approx 4.3\epsilon_T,$ was chosen as to minimize $\epsilon.$}
    \label{fig:QV_vsTGateErrorRate}
\end{figure}

In order to take into account the effect of finite fidelity magic states, we
first find the gate decomposition in terms of $T$ gates. As discussed in
\autoref{sec:PhysError}, for a sequence of arbitrary $SU(4)$ gates we can
decompose each step into three CNOTs and nine single qubit rotations
(see~\autoref{fig:SU4CircuitDiagram}). Each random rotation can be implemented
to a precision $\epsilon_P$ using a $T$ gate count of
$-3\log_2(\epsilon_P)$~\cite{SU2TCount2016}. Therefore, the effective error
rate (per qubit per step) is
\begin{equation} 
    \epsilon_\text{eff} \approx 4.5\epsilon_P -4.5*3\log_2(\epsilon_P)\epsilon_T + \epsilon_L.
    \label{eqn:Error_QEC}
\end{equation}
% Plots of QV-$k$ for imperfect $T$ gates are shown in
% \autoref{fig:QV_vsTGateErrorRate}.

\subsection{Optimized Fault Tolerant Architecture}
So far we have been considering piecemeal various aspects that are associated
with accounting for the trade-off between overhead and performance allowed by
error correction. In this section we consider the problem of trying to estimate
the optimal performance taking into account error code distance, $T$ gate
compilation and imprecision, and the overhead needed to produce the magic
states to implement these gates. In order to think about the broader
architecture in a specific way, we follow the general ideas laid out in
Ref.~\cite{GameOfSurfaceCodes2019}. 

\begin{figure}[htpb]
    \centering
    \includegraphics[width=\columnwidth]{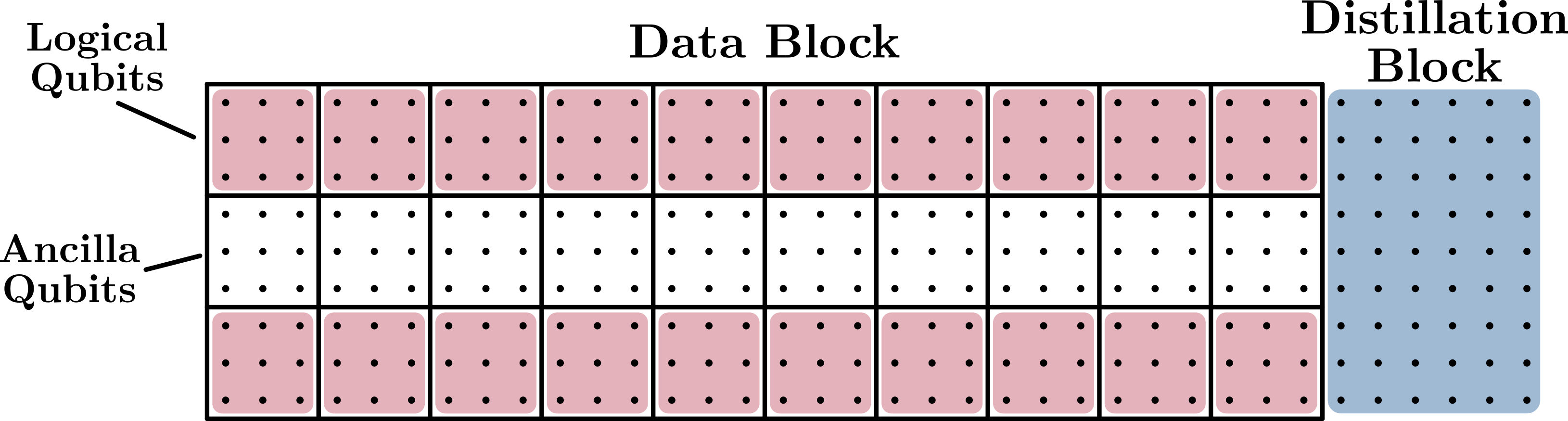}
    \caption{SurfaceCodeArchitecture}
    \label{fig:SurfaceCodeArchitecture}
\end{figure}

First, we assume that we have a given number of physical qubits. We logically
divide our qubits into two segments which we label the Data Block and the
Distillation Block (see \autoref{fig:SurfaceCodeArchitecture}). The Data Block
is further divided into blocks of qubits that will act as our logical qubits.
In addition, there will typically also need to be ancilla logical qubit blocks
in order to be able to move information around and/or account for the lack of
full connectivity in most large-scale architectures. We'll assume this overhead
to be 50\% of the Data Block, which is all that is needed for nearest neighbor
connectivity~\cite{GameOfSurfaceCodes2019}. 

\begin{figure}[htpb]
    \centering
    \includegraphics[width=\columnwidth]{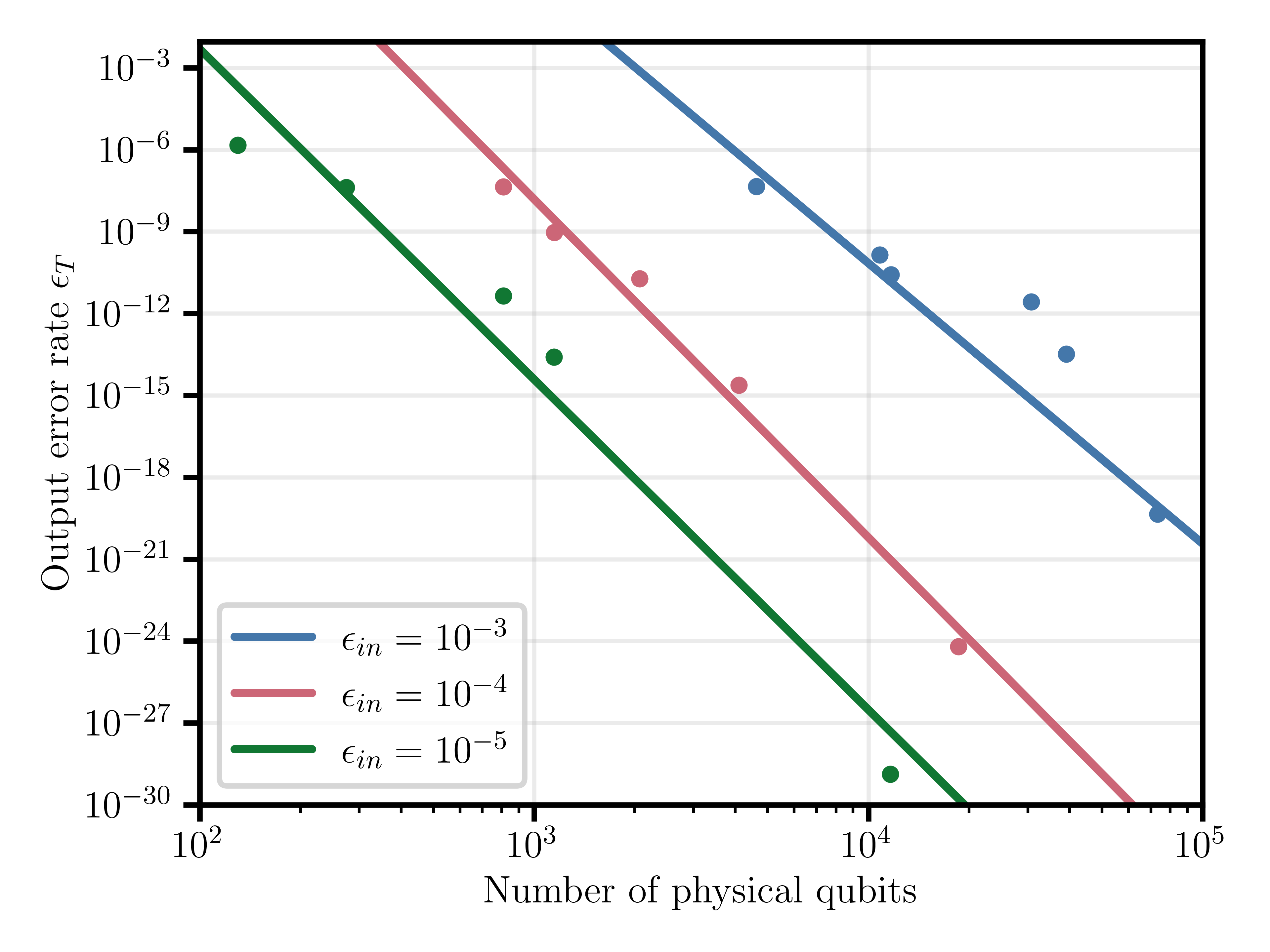}
    \caption{Magic state distillation overhead. The magic state error rate
        $\epsilon_T$ plotted as a function of the number of physical qubits in
        the Distillation Block for various physical error rates. Based upon the
        distillation scheme from Ref.~\cite{LitinskiMSD2019}.}
    \label{fig:MS_Overhead}
\end{figure}

The Distillation Block is the set of qubits designated to generating magic
states to be used in order to implement $T$ gates. The optimal strategy for
generating magic states is still an open area of research, but significant
progress has been made in bringing down the space-time
overhead~\cite{MagicStateDistillation2005, GidneyCCZMSD2019, LitinskiMSD2019}.
Generally, the distilled magic state fidelity is a function of the input
fidelity (which is roughly the same as the physical error
rate~\cite{MagicStateErrorPhysError2015}) and the overhead dedicated to
distillation. For this work we compute the overhead using the state-of-the art
optimized distillation schemes presented in Ref.~\cite{LitinskiMSD2019} and
optimizations were performed assuming a continuous functional trade off between
qubit number and output error rate as shown in \autoref{fig:MS_Overhead}. 

\begin{figure*}[htpb]
    \centering
    \includegraphics[width=\textwidth]{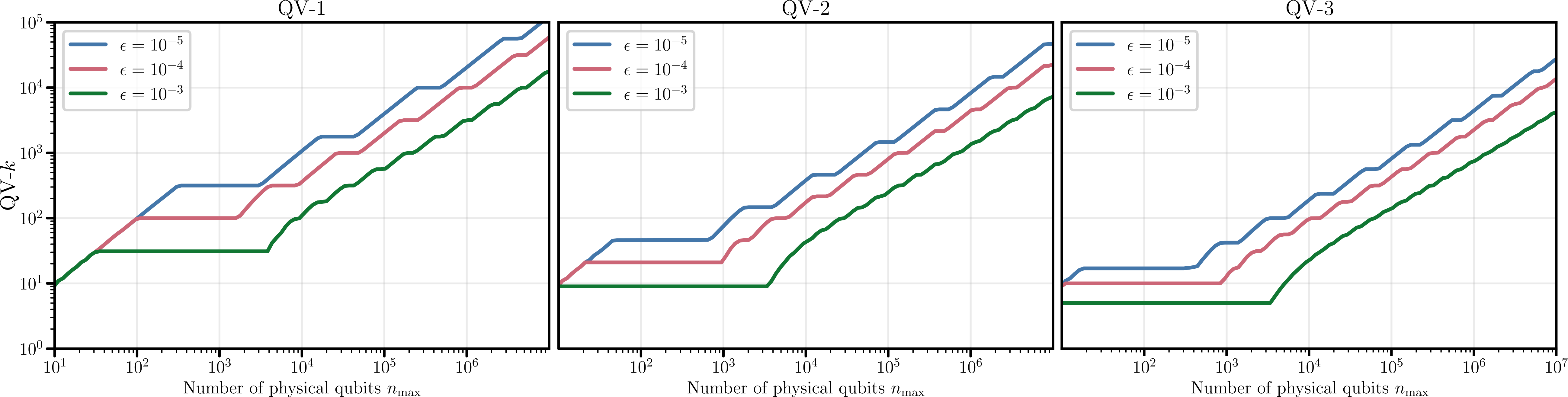}
    \caption{Optimization of the fault tolerant architecture including code
        distance, magic state distillation block size, and gate decomposition
        to maximize QV-$k$ for various values of the underlying physical error
        rate $\epsilon$.}
    \label{fig:QEC_QVk}
\end{figure*}

Combining all these pieces we can write QV Class as
\begin{equation}
    \text{QV-}k 
    % = \argmax_{1\le 1.5n_L + n_{MSD}\le n_\text{max}} 
    = \argmax_{n_D,\, d_c}
      \left(
          \min\left[n_L,\epsilon_\text{eff}^{-1/(k+1)}\right] 
      \right),
    \label{eqn:QVk_QEC}
\end{equation}
where $n_D$ is the number of qubits in the Distillation block and $d_c$ is the
code distance as discussed in \autoref{subsec:nLvsdc}. The number of logical
qubits $n_L$ now depends on both the code distance (as before) as well as
$n_D$, i.e. 
\begin{equation}
    n_L = \left\lfloor\frac{n_\text{max}-n_D}{1.5(2d_c-1)^2}\right\rfloor.
\end{equation}
The effective error rate $\epsilon_\text{eff}$ is given by
\autoref{eqn:Error_QEC} which depends upon both the logical error rate
$\epsilon_L$ (\autoref{eqn:err_LSC}) as well as the $T$ gate error rate and
thus the Distillation Block size $n_D$.

With these assumptions, we numerically solved \autoref{eqn:QVk_QEC} as a
function of total physical qubit number $n_\text{max}$ for a variety of
physical error rates. The results are presented in \autoref{fig:QEC_QVk}. Just
as in our naive QEC model in \autoref{subsec:nLvsdc}, initially there is a
continuous rise in the system performance as we add more qubits until we reach
the first initial plateau representing the regime in which we are limited by
the physical error rate but don't have enough qubits to afford the overhead
necessary to improve performance through error correction. However, due to the
relatively high cost of switching from physical to fault tolerant gates due to
the requirements of magic state generation, this plateau makes QEC only useful
once we reach a few thousand physical qubits. We note that the exact threshold
is sensitive to the efficiency of magic state generation and thus will improve
if/when better schemes are developed.

% flatex input end: [qec.tex]

%% Paper sections
% flatex input: [conclusion.tex]
\section{Conclusions}
Quantum computing is a rapidly developing technology. As this technology
matures, it is important to develop and use quantitative metrics to track and
communicate performance. Such performance metrics need to be useful and
understandable to potential end users if we want to see widespread adoption.
This means that we need quantum computing metrics that are connected to
applications. 

In our previous work, we showed how to create a set of metrics inspired by the
quantum volume that are better connected to end user
applications~\cite{DHS_paper1}. In the current work, we presented how these
metrics can be estimated given the knowledge of various system parameters.
These considerations included qubit number and connectivity of the device. We
also looked at effective error rates to implement computational primitives and
how these rates can be connected to the underlying physical gate set of the
device. 

Finally, we presented a sketch of how the trade-offs between resources and
error rates in the fault tolerant regime can be incorporated into estimating
these metrics. This optimization included the overhead needed to encode each
logical qubit, ancilla qubits for effective connectivity, errors related to
approximate gate decomposition, imperfect $T$ gates, and the overhead needed
for magic state generation. A key result is the estimation of thresholds in
terms of qubit numbers needed for a quantum error correction to provide
additional computational improvements beyond what is possible without error
correction on the same device. In particular, a quantum computer with a
physical error rate of $\epsilon=10^{-3}$ (slightly better than state of the
art) would need about 5,000 physical qubits before reaching this threshold. 

% flatex input end: [conclusion.tex]

%% Paper sections

% \section*{Acknowledgements}
% Add if we need to add an acknowledgements statement (e.g. DHS statement)

%% Bibliography
% \bibliographystyle{plainnat}
%*flatex input: [main.bbl]

% flatex input end: [main.bbl]
%FLATEX-REM:\bibliographystyle{unsrtnat}
%FLATEX-REM:\bibliography{References.bib}

\begin{thebibliography}{40}
\providecommand{\natexlab}[1]{#1}
\providecommand{\url}[1]{\texttt{#1}}
\expandafter\ifx\csname urlstyle\endcsname\relax
  \providecommand{\doi}[1]{doi: #1}\else
  \providecommand{\doi}{doi: \begingroup \urlstyle{rm}\Url}\fi

\bibitem[Miller et~al.(2022)Miller, Broomfield, Cox, Kinast, and
  Rodenburg]{DHS_paper1}
Keith Miller, Charles Broomfield, Ann Cox, Joe Kinast, and Brandon Rodenburg.
\newblock An {{Improved Volumetric Metric}} for {{Quantum Computers}} via more
  {{Representative Quantum Circuit Shapes}}.
\newblock \emph{arXiv:2207.02315 [quant-ph]}, 2022.
\newblock \doi{10.48550/arXiv.2207.02315}.

\bibitem[MacQuarrie et~al.(2020)MacQuarrie, Simon, Simmons, and
  Maine]{QCCommercial2020}
Evan~R. MacQuarrie, Christoph Simon, Stephanie Simmons, and Elicia Maine.
\newblock The emerging commercial landscape of quantum computing.
\newblock \emph{Nature Reviews Physics}, 2\penalty0 (11):\penalty0 596--598,
  2020.
\newblock \doi{10.1038/s42254-020-00247-5}.

\bibitem[Zap(2022)]{ZapataEnterpriseAdoption2022}
\emph{The {{Second Annual Report}} on {{Enterprise Quantum Computing
  Adoption}}}.
\newblock {Zapata}, December 2022.

\bibitem[Moll et~al.(2018)Moll, Barkoutsos, Bishop, Chow, Cross, Egger, {Stefan
  Filipp}, Fuhrer, Gambetta, Ganzhorn, Kandala, Mezzacapo, {Peter Müller},
  Riess, Salis, Smolin, Tavernelli, and Temme]{QV2018}
Nikolaj Moll, Panagiotis Barkoutsos, Lev~S. Bishop, Jerry~M. Chow, Andrew
  Cross, Daniel~J. Egger, {Stefan Filipp}, Andreas Fuhrer, Jay~M. Gambetta,
  Marc Ganzhorn, Abhinav Kandala, Antonio Mezzacapo, {Peter Müller}, Walter
  Riess, Gian Salis, John Smolin, Ivano Tavernelli, and Kristan Temme.
\newblock Quantum optimization using variational algorithms on near-term
  quantum devices.
\newblock \emph{Quantum Science and Technology}, 3\penalty0 (3):\penalty0
  030503, 2018.
\newblock \doi{10.1088/2058-9565/aab822}.

\bibitem[Cross et~al.(2019)Cross, Bishop, Sheldon, Nation, and
  Gambetta]{QV2019}
Andrew~W. Cross, Lev~S. Bishop, Sarah Sheldon, Paul~D. Nation, and Jay~M.
  Gambetta.
\newblock Validating quantum computers using randomized model circuits.
\newblock \emph{Physical Review A}, 100\penalty0 (3):\penalty0 032328, 2019.
\newblock \doi{10.1103/PhysRevA.100.032328}.

\bibitem[Sekino and Susskind(2008)]{FastScramble2008}
Yasuhiro Sekino and L.~Susskind.
\newblock Fast scramblers.
\newblock \emph{Journal of High Energy Physics}, 2008\penalty0 (10):\penalty0
  065--065, 2008.
\newblock \doi{10.1088/1126-6708/2008/10/065}.

\bibitem[Blume-Kohout and Young(2020)]{QVolumetricMetrics2020}
Robin Blume-Kohout and Kevin~C. Young.
\newblock A volumetric framework for quantum computer benchmarks.
\newblock \emph{Quantum}, 4:\penalty0 362, 2020.
\newblock \doi{10.22331/q-2020-11-15-362}.

\bibitem[Proctor et~al.(2021)Proctor, Rudinger, Young, Nielsen, and
  {Blume-Kohout}]{MirrorCircuitBenchmarking2021}
Timothy Proctor, Kenneth Rudinger, Kevin Young, Erik Nielsen, and Robin
  {Blume-Kohout}.
\newblock Measuring the capabilities of quantum computers.
\newblock \emph{Nature Physics}, pages 1--5, 2021.
\newblock \doi{10.1038/s41567-021-01409-7}.

\bibitem[Lubinski et~al.(2023)Lubinski, Johri, Varosy, Coleman, Zhao, Necaise,
  Baldwin, Mayer, and Proctor]{QEDC_ApplicationBenchmarks2021}
Thomas Lubinski, Sonika Johri, Paul Varosy, Jeremiah Coleman, Luning Zhao,
  Jason Necaise, Charles~H. Baldwin, Karl Mayer, and Timothy Proctor.
\newblock Application-{{Oriented Performance Benchmarks}} for {{Quantum
  Computing}}.
\newblock \emph{IEEE Transactions on Quantum Engineering}, 4:\penalty0 1--32,
  2023.
\newblock ISSN 2689-1808.
\newblock \doi{10.1109/TQE.2023.3253761}.

\bibitem[Hamilton et~al.(2022)Hamilton, Laanait, Francis, Economou, Barron,
  {Yeter-Aydeniz}, Morris, Cooley, Kang, Kemper, and
  Pooser]{EntanglementBenchmark2022}
Kathleen~E. Hamilton, Nouamane Laanait, Akhil Francis, Sophia~E. Economou,
  George~S. Barron, K{\"u}bra {Yeter-Aydeniz}, Titus Morris, Harrison Cooley,
  Muhun Kang, Alexander~F. Kemper, and Raphael Pooser.
\newblock An entanglement-based volumetric benchmark for near-term quantum
  hardware.
\newblock \emph{arXiv:2209.00678}, 2022.
\newblock \doi{10.48550/arXiv.2209.00678}.

\bibitem[Shor(1994)]{ShorAlg1994}
P.~W. Shor.
\newblock Algorithms for quantum computation: Discrete logarithms and
  factoring.
\newblock In \emph{Proceedings 35th {{Annual Symposium}} on {{Foundations}} of
  {{Computer Science}}}, pages 124--134, 1994.
\newblock \doi{10.1109/SFCS.1994.365700}.

\bibitem[Brown et~al.(2016)Brown, Kim, and Monroe]{IonQCNetworkTopology2016}
Kenneth~R. Brown, Jungsang Kim, and Christopher Monroe.
\newblock Co-designing a scalable quantum computer with trapped atomic ions.
\newblock \emph{npj Quantum Information}, 2:\penalty0 16034, 2016.
\newblock ISSN 2056-6387.
\newblock \doi{10.1038/npjqi.2016.34}.

\bibitem[Knill et~al.(2008)Knill, Leibfried, Reichle, Britton, Blakestad, Jost,
  Langer, Ozeri, Seidelin, and Wineland]{RandomizedBenchmarking2008}
E.~Knill, D.~Leibfried, R.~Reichle, J.~Britton, R.~B. Blakestad, J.~D. Jost,
  C.~Langer, R.~Ozeri, S.~Seidelin, and D.~J. Wineland.
\newblock Randomized {{Benchmarking}} of {{Quantum Gates}}.
\newblock \emph{Physical Review A}, 77\penalty0 (1):\penalty0 012307, 2008.
\newblock ISSN 1050-2947, 1094-1622.
\newblock \doi{10.1103/PhysRevA.77.012307}.

\bibitem[Proctor et~al.(2022)Proctor, Rudinger, Young, Nielsen, and
  {Blume-Kohout}]{MirrorCircuitBenchmarking2022}
Timothy Proctor, Kenneth Rudinger, Kevin Young, Erik Nielsen, and Robin
  {Blume-Kohout}.
\newblock Measuring the {{Capabilities}} of {{Quantum Computers}}.
\newblock \emph{Nature Physics}, 18\penalty0 (1):\penalty0 75--79, 2022.
\newblock ISSN 1745-2473, 1745-2481.
\newblock \doi{10.1038/s41567-021-01409-7}.

\bibitem[Wallman and Emerson(2016)]{RandomizedCompiling2016}
Joel~J. Wallman and Joseph Emerson.
\newblock Noise tailoring for scalable quantum computation via randomized
  compiling.
\newblock \emph{Physical Review A}, 94\penalty0 (5):\penalty0 052325, 2016.
\newblock ISSN 2469-9926, 2469-9934.
\newblock \doi{10.1103/PhysRevA.94.052325}.

\bibitem[Hashim et~al.(2021)Hashim, Naik, Morvan, Ville, Mitchell, Kreikebaum,
  Davis, Smith, Iancu, O'Brien, Hincks, Wallman, Emerson, and
  Siddiqi]{RandomizedCompiling2021}
Akel Hashim, Ravi~K. Naik, Alexis Morvan, Jean-Loup Ville, Bradley Mitchell,
  John~Mark Kreikebaum, Marc Davis, Ethan Smith, Costin Iancu, Kevin~P.
  O'Brien, Ian Hincks, Joel~J. Wallman, Joseph Emerson, and Irfan Siddiqi.
\newblock Randomized compiling for scalable quantum computing on a noisy
  superconducting quantum processor.
\newblock \emph{Physical Review X}, 11\penalty0 (4):\penalty0 041039, 2021.
\newblock ISSN 2160-3308.
\newblock \doi{10.1103/PhysRevX.11.041039}.

\bibitem[Vatan and Williams(2004)]{Optimal2QbGates2004}
Farrokh Vatan and Colin Williams.
\newblock Optimal {{Quantum Circuits}} for {{General Two-Qubit Gates}}.
\newblock \emph{Physical Review A}, 69\penalty0 (3):\penalty0 032315, 2004.
\newblock ISSN 1050-2947, 1094-1622.
\newblock \doi{10.1103/PhysRevA.69.032315}.

\bibitem[Tomesh et~al.(2022)Tomesh, Gokhale, Omole, Ravi, Smith, Viszlai, Wu,
  Hardavellas, Martonosi, and Chong]{SupermarQBehcnmark2022}
Teague Tomesh, Pranav Gokhale, Victory Omole, Gokul~Subramanian Ravi,
  Kaitlin~N. Smith, Joshua Viszlai, Xin-Chuan Wu, Nikos Hardavellas,
  Margaret~R. Martonosi, and Frederic~T. Chong.
\newblock {{SupermarQ}}: {{A Scalable Quantum Benchmark Suite}}.
\newblock In \emph{2022 {{IEEE International Symposium}} on {{High-Performance
  Computer Architecture}} ({{HPCA}})}, pages 587--603, 2022.
\newblock \doi{10.1109/HPCA53966.2022.00050}.

\bibitem[Harper and Flammia(2019)]{QEC1}
Robin Harper and Steven~T. Flammia.
\newblock Fault-{Tolerant} {Logical} {Gates} in the {IBM} {Quantum}
  {Experience}.
\newblock \emph{Physical Review Letters}, 122\penalty0 (8):\penalty0 080504,
  2019.
\newblock \doi{10.1103/PhysRevLett.122.080504}.

\bibitem[{Google Quantum AI} et~al.(2021){Google Quantum AI}, Chen, Satzinger,
  Atalaya, Korotkov, Dunsworth, Sank, Quintana, McEwen, Barends, Klimov, Hong,
  Jones, Petukhov, Kafri, Demura, Burkett, Gidney, Fowler, Paler, Putterman,
  Aleiner, Arute, Arya, Babbush, Bardin, Bengtsson, Bourassa, Broughton,
  Buckley, Buell, Bushnell, Chiaro, Collins, Courtney, Derk, Eppens, Erickson,
  Farhi, Foxen, Giustina, Greene, Gross, Harrigan, Harrington, Hilton, Ho,
  Huang, Huggins, Ioffe, Isakov, Jeffrey, Jiang, Kechedzhi, Kim, Kitaev,
  Kostritsa, Landhuis, Laptev, Lucero, Martin, McClean, McCourt, Mi, Miao,
  Mohseni, Montazeri, Mruczkiewicz, Mutus, Naaman, Neeley, Neill, Newman, Niu,
  O’Brien, Opremcak, Ostby, Pató, Redd, Roushan, Rubin, Shvarts, Strain,
  Szalay, Trevithick, Villalonga, White, Yao, Yeh, Yoo, Zalcman, Neven, Boixo,
  Smelyanskiy, Chen, Megrant, and Kelly]{QEC2}
{Google Quantum AI}, Zijun Chen, Kevin~J. Satzinger, Juan Atalaya, Alexander~N.
  Korotkov, Andrew Dunsworth, Daniel Sank, Chris Quintana, Matt McEwen, Rami
  Barends, Paul~V. Klimov, Sabrina Hong, Cody Jones, Andre Petukhov, Dvir
  Kafri, Sean Demura, Brian Burkett, Craig Gidney, Austin~G. Fowler, Alexandru
  Paler, Harald Putterman, Igor Aleiner, Frank Arute, Kunal Arya, Ryan Babbush,
  Joseph~C. Bardin, Andreas Bengtsson, Alexandre Bourassa, Michael Broughton,
  Bob~B. Buckley, David~A. Buell, Nicholas Bushnell, Benjamin Chiaro, Roberto
  Collins, William Courtney, Alan~R. Derk, Daniel Eppens, Catherine Erickson,
  Edward Farhi, Brooks Foxen, Marissa Giustina, Ami Greene, Jonathan~A. Gross,
  Matthew~P. Harrigan, Sean~D. Harrington, Jeremy Hilton, Alan Ho, Trent Huang,
  William~J. Huggins, L.~B. Ioffe, Sergei~V. Isakov, Evan Jeffrey, Zhang Jiang,
  Kostyantyn Kechedzhi, Seon Kim, Alexei Kitaev, Fedor Kostritsa, David
  Landhuis, Pavel Laptev, Erik Lucero, Orion Martin, Jarrod~R. McClean, Trevor
  McCourt, Xiao Mi, Kevin~C. Miao, Masoud Mohseni, Shirin Montazeri, Wojciech
  Mruczkiewicz, Josh Mutus, Ofer Naaman, Matthew Neeley, Charles Neill, Michael
  Newman, Murphy~Yuezhen Niu, Thomas~E. O’Brien, Alex Opremcak, Eric Ostby,
  Bálint Pató, Nicholas Redd, Pedram Roushan, Nicholas~C. Rubin, Vladimir
  Shvarts, Doug Strain, Marco Szalay, Matthew~D. Trevithick, Benjamin
  Villalonga, Theodore White, Z.~Jamie Yao, Ping Yeh, Juhwan Yoo, Adam Zalcman,
  Hartmut Neven, Sergio Boixo, Vadim Smelyanskiy, Yu~Chen, Anthony Megrant, and
  Julian Kelly.
\newblock Exponential suppression of bit or phase errors with cyclic error
  correction.
\newblock \emph{Nature}, 595\penalty0 (7867):\penalty0 383--387, 2021.
\newblock \doi{10.1038/s41586-021-03588-y}.

\bibitem[Erhard et~al.(2021)Erhard, Poulsen~Nautrup, Meth, Postler, Stricker,
  Stadler, Negnevitsky, Ringbauer, Schindler, Briegel, Blatt, Friis, and
  Monz]{QEC3}
Alexander Erhard, Hendrik Poulsen~Nautrup, Michael Meth, Lukas Postler, Roman
  Stricker, Martin Stadler, Vlad Negnevitsky, Martin Ringbauer, Philipp
  Schindler, Hans~J. Briegel, Rainer Blatt, Nicolai Friis, and Thomas Monz.
\newblock Entangling logical qubits with lattice surgery.
\newblock \emph{Nature}, 589\penalty0 (7841):\penalty0 220--224, 2021.
\newblock \doi{10.1038/s41586-020-03079-6}.

\bibitem[Vigliar et~al.(2021)Vigliar, Paesani, Ding, Adcock, Wang,
  Morley-Short, Bacco, Oxenløwe, Thompson, Rarity, and Laing]{QEC4}
Caterina Vigliar, Stefano Paesani, Yunhong Ding, Jeremy~C. Adcock, Jianwei
  Wang, Sam Morley-Short, Davide Bacco, Leif~K. Oxenløwe, Mark~G. Thompson,
  John~G. Rarity, and Anthony Laing.
\newblock Error-protected qubits in a silicon photonic chip.
\newblock \emph{Nature Physics}, pages 1--7, 2021.
\newblock \doi{10.1038/s41567-021-01333-w}.

\bibitem[Egan et~al.(2021)Egan, Debroy, Noel, Risinger, Zhu, Biswas, Newman,
  Li, Brown, Cetina, and Monroe]{QEC5}
Laird Egan, Dripto~M. Debroy, Crystal Noel, Andrew Risinger, Daiwei Zhu,
  Debopriyo Biswas, Michael Newman, Muyuan Li, Kenneth~R. Brown, Marko Cetina,
  and Christopher Monroe.
\newblock Fault-tolerant control of an error-corrected qubit.
\newblock \emph{Nature}, pages 1--6, 2021.
\newblock \doi{10.1038/s41586-021-03928-y}.

\bibitem[{Ryan-Anderson} et~al.(2021){Ryan-Anderson}, Bohnet, Lee, Gresh,
  Hankin, Gaebler, Francois, Chernoguzov, Lucchetti, Brown, Gatterman, Halit,
  Gilmore, Gerber, Neyenhuis, Hayes, and Stutz]{QEC6}
C.~{Ryan-Anderson}, J.~G. Bohnet, K.~Lee, D.~Gresh, A.~Hankin, J.~P. Gaebler,
  D.~Francois, A.~Chernoguzov, D.~Lucchetti, N.~C. Brown, T.~M. Gatterman,
  S.~K. Halit, K.~Gilmore, J.~A. Gerber, B.~Neyenhuis, D.~Hayes, and R.~P.
  Stutz.
\newblock Realization of {Real}-{Time} {Fault}-{Tolerant} {Quantum} {Error}
  {Correction}.
\newblock \emph{Physical Review X}, 11\penalty0 (4):\penalty0 041058, 2021.
\newblock \doi{10.1103/PhysRevX.11.041058}.

\bibitem[Postler et~al.(2022)Postler, Heu{\ss}en, Pogorelov, Rispler, Feldker,
  Meth, Marciniak, Stricker, Ringbauer, Blatt, Schindler, M{\"u}ller, and
  Monz]{QEC7}
Lukas Postler, Sascha Heu{\ss}en, Ivan Pogorelov, Manuel Rispler, Thomas
  Feldker, Michael Meth, Christian~D. Marciniak, Roman Stricker, Martin
  Ringbauer, Rainer Blatt, Philipp Schindler, Markus M{\"u}ller, and Thomas
  Monz.
\newblock Demonstration of fault-tolerant universal quantum gate operations.
\newblock \emph{Nature}, 605\penalty0 (7911):\penalty0 675--680, 2022.
\newblock \doi{10.1038/s41586-022-04721-1}.

\bibitem[Zhao et~al.(2022)Zhao, Ye, Huang, Zhang, Wu, Guan, Zhu, Wei, He, Cao,
  Chen, Chung, Deng, Fan, Gong, Guo, Guo, Han, Li, Li, Li, Liang, Lin, Qian,
  Rong, Su, Sun, Wang, Wu, Xu, Ying, Yu, Zha, Zhang, Huo, Lu, Peng, Zhu, and
  Pan]{QEC8}
Youwei Zhao, Yangsen Ye, He-Liang Huang, Yiming Zhang, Dachao Wu, Huijie Guan,
  Qingling Zhu, Zuolin Wei, Tan He, Sirui Cao, Fusheng Chen, Tung-Hsun Chung,
  Hui Deng, Daojin Fan, Ming Gong, Cheng Guo, Shaojun Guo, Lianchen Han, Na~Li,
  Shaowei Li, Yuan Li, Futian Liang, Jin Lin, Haoran Qian, Hao Rong, Hong Su,
  Lihua Sun, Shiyu Wang, Yulin Wu, Yu~Xu, Chong Ying, Jiale Yu, Chen Zha, Kaili
  Zhang, Yong-Heng Huo, Chao-Yang Lu, Cheng-Zhi Peng, Xiaobo Zhu, and Jian-Wei
  Pan.
\newblock Realization of an {{Error-Correcting Surface Code}} with
  {{Superconducting Qubits}}.
\newblock \emph{Physical Review Letters}, 129\penalty0 (3):\penalty0 030501,
  2022.
\newblock \doi{10.1103/PhysRevLett.129.030501}.

\bibitem[Sivak et~al.(2023)Sivak, Eickbusch, Royer, Singh, Tsioutsios, Ganjam,
  Miano, Brock, Ding, Frunzio, Girvin, Schoelkopf, and Devoret]{QEC9}
V.~V. Sivak, A.~Eickbusch, B.~Royer, S.~Singh, I.~Tsioutsios, S.~Ganjam,
  A.~Miano, B.~L. Brock, A.~Z. Ding, L.~Frunzio, S.~M. Girvin, R.~J.
  Schoelkopf, and M.~H. Devoret.
\newblock Real-time quantum error correction beyond break-even.
\newblock \emph{Nature}, 616\penalty0 (7955):\penalty0 50--55, April 2023.
\newblock ISSN 1476-4687.
\newblock \doi{10.1038/s41586-023-05782-6}.

\bibitem[Reichardt et~al.(2024)Reichardt, Paetznick, Aasen, Basov,
  {Bello-Rivas}, Bonderson, Chao, van Dam, Hastings, Paz, da~Silva, Sundaram,
  Svore, Vaschillo, Wang, Zanner, Cairncross, Chen, Crow, Kim, Kindem, King,
  McDonald, Norcia, Ryou, Stone, Wadleigh, Barnes, Battaglino, Bohdanowicz,
  Booth, Brown, Brown, Cassella, Coxe, Epstein, Feldkamp, Griger, Halperin,
  Heinz, Hummel, Jaffe, Jones, Kapit, Kotru, Lauigan, Li, Marjanovic, Megidish,
  Meredith, Morshead, Muniz, Narayanaswami, Nishiguchi, Paule, Pawlak, Pudenz,
  P{\'e}rez, Simon, Smull, Stack, Urbanek, van~de Veerdonk, Vendeiro, Weverka,
  Wilkason, Wu, Xie, {Zalys-Geller}, Zhang, and Bloom]{QEC10}
Ben~W. Reichardt, Adam Paetznick, David Aasen, Ivan Basov, Juan~M.
  {Bello-Rivas}, Parsa Bonderson, Rui Chao, Wim van Dam, Matthew~B. Hastings,
  Andres Paz, Marcus~P. da~Silva, Aarthi Sundaram, Krysta~M. Svore, Alexander
  Vaschillo, Zhenghan Wang, Matt Zanner, William~B. Cairncross, Cheng-An Chen,
  Daniel Crow, Hyosub Kim, Jonathan~M. Kindem, Jonathan King, Michael McDonald,
  Matthew~A. Norcia, Albert Ryou, Mark Stone, Laura Wadleigh, Katrina Barnes,
  Peter Battaglino, Thomas~C. Bohdanowicz, Graham Booth, Andrew Brown, Mark~O.
  Brown, Kayleigh Cassella, Robin Coxe, Jeffrey~M. Epstein, Max Feldkamp,
  Christopher Griger, Eli Halperin, Andre Heinz, Frederic Hummel, Matthew
  Jaffe, Antonia M.~W. Jones, Eliot Kapit, Krish Kotru, Joseph Lauigan, Ming
  Li, Jan Marjanovic, Eli Megidish, Matthew Meredith, Ryan Morshead, Juan~A.
  Muniz, Sandeep Narayanaswami, Ciro Nishiguchi, Timothy Paule, Kelly~A.
  Pawlak, Kristen~L. Pudenz, David~Rodr{\'i}guez P{\'e}rez, Jon Simon, Aaron
  Smull, Daniel Stack, Miroslav Urbanek, Ren{\'e} J.~M. van~de Veerdonk,
  Zachary Vendeiro, Robert~T. Weverka, Thomas Wilkason, Tsung-Yao Wu, Xin Xie,
  Evan {Zalys-Geller}, Xiaogang Zhang, and Benjamin~J. Bloom.
\newblock Logical computation demonstrated with a neutral atom quantum
  processor.
\newblock \emph{arXiv:2411.11822 [quant-ph]}, 2024.
\newblock \doi{10.48550/arXiv.2411.11822}.

\bibitem[Acharya et~al.(2024)Acharya, Abanin, {Aghababaie-Beni}, Aleiner,
  Andersen, Ansmann, Arute, Arya, Asfaw, Astrakhantsev, Atalaya, Babbush,
  Bacon, Ballard, Bardin, Bausch, Bengtsson, Bilmes, Blackwell, Boixo, Bortoli,
  Bourassa, Bovaird, Brill, Broughton, Browne, Buchea, Buckley, Buell, Burger,
  Burkett, Bushnell, Cabrera, Campero, Chang, Chen, Chen, Chiaro, Chik, Chou,
  Claes, Cleland, Cogan, Collins, Conner, Courtney, Crook, Curtin, Das, Davies,
  De~Lorenzo, Debroy, Demura, Devoret, Di~Paolo, Donohoe, Drozdov, Dunsworth,
  Earle, Edlich, Eickbusch, Elbag, Elzouka, Erickson, Faoro, Farhi, Ferreira,
  Burgos, Forati, Fowler, Foxen, Ganjam, Garcia, Gasca, Genois, Giang, Gidney,
  Gilboa, Gosula, Dau, Graumann, Greene, Gross, Habegger, Hall, Hamilton,
  Hansen, Harrigan, Harrington, Heras, Heslin, Heu, Higgott, Hill, Hilton,
  Holland, Hong, Huang, Huff, Huggins, Ioffe, Isakov, Iveland, Jeffrey, Jiang,
  Jones, Jordan, Joshi, Juhas, Kafri, Kang, Karamlou, Kechedzhi, Kelly, Khaire,
  Khattar, Khezri, Kim, Klimov, Klots, Kobrin, Kohli, Korotkov, Kostritsa,
  Kothari, Kozlovskii, Kreikebaum, Kurilovich, Lacroix, Landhuis, {Lange-Dei},
  Langley, Laptev, Lau, Le~Guevel, Ledford, Lee, Lee, Lensky, Leon, Lester, Li,
  Li, Lill, Liu, Livingston, Locharla, Lucero, Lundahl, Lunt, Madhuk, Malone,
  Maloney, Mandr{\`a}, Manyika, Martin, Martin, Martin, Maxfield, McClean,
  McEwen, Meeks, Megrant, Mi, Miao, Mieszala, Molavi, Molina, Montazeri,
  Morvan, Movassagh, Mruczkiewicz, Naaman, Neeley, Neill, Nersisyan, Neven,
  Newman, Ng, Nguyen, Nguyen, Ni, Niu, O'Brien, Oliver, Opremcak, Ottosson,
  Petukhov, Pizzuto, Platt, Potter, Pritchard, Pryadko, Quintana, Ramachandran,
  Reagor, Redding, Rhodes, Roberts, Rosenberg, Rosenfeld, Roushan, Rubin, Saei,
  Sank, Sankaragomathi, Satzinger, Schurkus, Schuster, Senior, Shearn, Shorter,
  Shutty, Shvarts, Singh, Sivak, Skruzny, Small, Smelyanskiy, Smith, Somma,
  Springer, Sterling, Strain, Suchard, Szasz, Sztein, Thor, Torres, Torunbalci,
  Vaishnav, Vargas, Vdovichev, Vidal, Villalonga, Heidweiller, Waltman, Wang,
  Ware, Weber, Weidel, White, Wong, Woo, Xing, Yao, Yeh, Ying, Yoo, Yosri,
  Young, Zalcman, Zhang, Zhu, Zobrist, and {Google Quantum AI and
  Collaborators}]{QEC11}
Rajeev Acharya, Dmitry~A. Abanin, Laleh {Aghababaie-Beni}, Igor Aleiner,
  Trond~I. Andersen, Markus Ansmann, Frank Arute, Kunal Arya, Abraham Asfaw,
  Nikita Astrakhantsev, Juan Atalaya, Ryan Babbush, Dave Bacon, Brian Ballard,
  Joseph~C. Bardin, Johannes Bausch, Andreas Bengtsson, Alexander Bilmes, Sam
  Blackwell, Sergio Boixo, Gina Bortoli, Alexandre Bourassa, Jenna Bovaird,
  Leon Brill, Michael Broughton, David~A. Browne, Brett Buchea, Bob~B. Buckley,
  David~A. Buell, Tim Burger, Brian Burkett, Nicholas Bushnell, Anthony
  Cabrera, Juan Campero, Hung-Shen Chang, Yu~Chen, Zijun Chen, Ben Chiaro,
  Desmond Chik, Charina Chou, Jahan Claes, Agnetta~Y. Cleland, Josh Cogan,
  Roberto Collins, Paul Conner, William Courtney, Alexander~L. Crook, Ben
  Curtin, Sayan Das, Alex Davies, Laura De~Lorenzo, Dripto~M. Debroy, Sean
  Demura, Michel Devoret, Agustin Di~Paolo, Paul Donohoe, Ilya Drozdov, Andrew
  Dunsworth, Clint Earle, Thomas Edlich, Alec Eickbusch, Aviv~Moshe Elbag,
  Mahmoud Elzouka, Catherine Erickson, Lara Faoro, Edward Farhi, Vinicius~S.
  Ferreira, Leslie~Flores Burgos, Ebrahim Forati, Austin~G. Fowler, Brooks
  Foxen, Suhas Ganjam, Gonzalo Garcia, Robert Gasca, {\'E}lie Genois, William
  Giang, Craig Gidney, Dar Gilboa, Raja Gosula, Alejandro~Grajales Dau,
  Dietrich Graumann, Alex Greene, Jonathan~A. Gross, Steve Habegger, John Hall,
  Michael~C. Hamilton, Monica Hansen, Matthew~P. Harrigan, Sean~D. Harrington,
  Francisco J.~H. Heras, Stephen Heslin, Paula Heu, Oscar Higgott, Gordon Hill,
  Jeremy Hilton, George Holland, Sabrina Hong, Hsin-Yuan Huang, Ashley Huff,
  William~J. Huggins, Lev~B. Ioffe, Sergei~V. Isakov, Justin Iveland, Evan
  Jeffrey, Zhang Jiang, Cody Jones, Stephen Jordan, Chaitali Joshi, Pavol
  Juhas, Dvir Kafri, Hui Kang, Amir~H. Karamlou, Kostyantyn Kechedzhi, Julian
  Kelly, Trupti Khaire, Tanuj Khattar, Mostafa Khezri, Seon Kim, Paul~V.
  Klimov, Andrey~R. Klots, Bryce Kobrin, Pushmeet Kohli, Alexander~N. Korotkov,
  Fedor Kostritsa, Robin Kothari, Borislav Kozlovskii, John~Mark Kreikebaum,
  Vladislav~D. Kurilovich, Nathan Lacroix, David Landhuis, Tiano {Lange-Dei},
  Brandon~W. Langley, Pavel Laptev, Kim-Ming Lau, Lo{\"i}ck Le~Guevel, Justin
  Ledford, Joonho Lee, Kenny Lee, Yuri~D. Lensky, Shannon Leon, Brian~J.
  Lester, Wing~Yan Li, Yin Li, Alexander~T. Lill, Wayne Liu, William~P.
  Livingston, Aditya Locharla, Erik Lucero, Daniel Lundahl, Aaron Lunt, Sid
  Madhuk, Fionn~D. Malone, Ashley Maloney, Salvatore Mandr{\`a}, James Manyika,
  Leigh~S. Martin, Orion Martin, Steven Martin, Cameron Maxfield, Jarrod~R.
  McClean, Matt McEwen, Seneca Meeks, Anthony Megrant, Xiao Mi, Kevin~C. Miao,
  Amanda Mieszala, Reza Molavi, Sebastian Molina, Shirin Montazeri, Alexis
  Morvan, Ramis Movassagh, Wojciech Mruczkiewicz, Ofer Naaman, Matthew Neeley,
  Charles Neill, Ani Nersisyan, Hartmut Neven, Michael Newman, Jiun~How Ng,
  Anthony Nguyen, Murray Nguyen, Chia-Hung Ni, Murphy~Yuezhen Niu, Thomas~E.
  O'Brien, William~D. Oliver, Alex Opremcak, Kristoffer Ottosson, Andre
  Petukhov, Alex Pizzuto, John Platt, Rebecca Potter, Orion Pritchard,
  Leonid~P. Pryadko, Chris Quintana, Ganesh Ramachandran, Matthew~J. Reagor,
  John Redding, David~M. Rhodes, Gabrielle Roberts, Eliott Rosenberg, Emma
  Rosenfeld, Pedram Roushan, Nicholas~C. Rubin, Negar Saei, Daniel Sank, Kannan
  Sankaragomathi, Kevin~J. Satzinger, Henry~F. Schurkus, Christopher Schuster,
  Andrew~W. Senior, Michael~J. Shearn, Aaron Shorter, Noah Shutty, Vladimir
  Shvarts, Shraddha Singh, Volodymyr Sivak, Jindra Skruzny, Spencer Small,
  Vadim Smelyanskiy, W.~Clarke Smith, Rolando~D. Somma, Sofia Springer, George
  Sterling, Doug Strain, Jordan Suchard, Aaron Szasz, Alex Sztein, Douglas
  Thor, Alfredo Torres, M.~Mert Torunbalci, Abeer Vaishnav, Justin Vargas,
  Sergey Vdovichev, Guifre Vidal, Benjamin Villalonga, Catherine~Vollgraff
  Heidweiller, Steven Waltman, Shannon~X. Wang, Brayden Ware, Kate Weber,
  Travis Weidel, Theodore White, Kristi Wong, Bryan W.~K. Woo, Cheng Xing,
  Z.~Jamie Yao, Ping Yeh, Bicheng Ying, Juhwan Yoo, Noureldin Yosri, Grayson
  Young, Adam Zalcman, Yaxing Zhang, Ningfeng Zhu, Nicholas Zobrist, and
  {Google Quantum AI and Collaborators}.
\newblock Quantum error correction below the surface code threshold.
\newblock \emph{Nature}, pages 1--3, December 2024.
\newblock ISSN 1476-4687.
\newblock \doi{10.1038/s41586-024-08449-y}.

\bibitem[Fowler et~al.(2012)Fowler, Mariantoni, Martinis, and
  Cleland]{QECSurfaceCode2012}
Austin~G. Fowler, Matteo Mariantoni, John~M. Martinis, and Andrew~N. Cleland.
\newblock Surface codes: {{Towards}} practical large-scale quantum computation.
\newblock \emph{Physical Review A}, 86\penalty0 (3):\penalty0 032324, 2012.
\newblock ISSN 1050-2947, 1094-1622.
\newblock \doi{10.1103/PhysRevA.86.032324}.

\bibitem[Sevilla and Riedel(2020)]{ForecastingQCTimelines2020}
Jaime Sevilla and C.~Jess Riedel.
\newblock Forecasting timelines of quantum computing.
\newblock \emph{arXiv:2009.05045 [quant-ph]}, 2020.
\newblock \doi{10.48550/arXiv.2009.05045}.

\bibitem[Eastin and Knill(2009)]{EastinKnillNoGo2009}
Bryan Eastin and Emanuel Knill.
\newblock Restrictions on {{Transversal Encoded Quantum Gate Sets}}.
\newblock \emph{Physical Review Letters}, 102\penalty0 (11):\penalty0 110502,
  2009.
\newblock ISSN 0031-9007, 1079-7114.
\newblock \doi{10.1103/PhysRevLett.102.110502}.

\bibitem[Gottesman(1998)]{GottesmanKnillTheorem1998}
Daniel Gottesman.
\newblock The {{Heisenberg Representation}} of {{Quantum Computers}}.
\newblock \emph{arXiv:quant-ph/9807006}, 1998.
\newblock \doi{10.48550/arXiv.quant-ph/9807006}.

\bibitem[Campbell et~al.(2017)Campbell, Terhal, and Vuillot]{FTQCReview2017}
Earl~T. Campbell, Barbara~M. Terhal, and Christophe Vuillot.
\newblock Roads towards fault-tolerant universal quantum computation.
\newblock \emph{Nature}, 549\penalty0 (7671):\penalty0 172--179, 2017.
\newblock ISSN 0028-0836.
\newblock \doi{10.1038/nature23460}.

\bibitem[Gidney and Fowler(2019)]{GidneyCCZMSD2019}
Craig Gidney and Austin~G. Fowler.
\newblock Efficient magic state factories with a catalyzed
  {\textbar}{{CCZ}}{$\rangle$} to 2 {\textbar}{{T}}{$\rangle$} transformation.
\newblock \emph{Quantum}, 3:\penalty0 135, 2019.
\newblock ISSN 2521-327X.
\newblock \doi{10.22331/q-2019-04-30-135}.

\bibitem[Bravyi and Kitaev(2005)]{MagicStateDistillation2005}
Sergei Bravyi and Alexei Kitaev.
\newblock Universal {{Quantum Computation}} with ideal {{Clifford}} gates and
  noisy ancillas.
\newblock \emph{Physical Review A}, 71\penalty0 (2), 2005.
\newblock ISSN 1050-2947, 1094-1622.
\newblock \doi{10.1103/PhysRevA.71.022316}.

\bibitem[Litinski(2019{\natexlab{a}})]{GameOfSurfaceCodes2019}
Daniel Litinski.
\newblock A {{Game}} of {{Surface Codes}}: {{Large-Scale Quantum Computing}}
  with {{Lattice Surgery}}.
\newblock \emph{Quantum}, 3:\penalty0 128, 2019{\natexlab{a}}.
\newblock \doi{10.22331/q-2019-03-05-128}.

\bibitem[Litinski(2019{\natexlab{b}})]{LitinskiMSD2019}
Daniel Litinski.
\newblock Magic {{State Distillation}}: {{Not}} as {{Costly}} as {{You Think}}.
\newblock \emph{Quantum}, 3:\penalty0 205, 2019{\natexlab{b}}.
\newblock \doi{10.22331/q-2019-12-02-205}.

\bibitem[Ross and Selinger(2016)]{SU2TCount2016}
Neil~J. Ross and Peter Selinger.
\newblock Optimal ancilla-free {{Clifford}}+{{T}} approximation of z-rotations,
  2016.

\bibitem[Li(2015)]{MagicStateErrorPhysError2015}
Ying Li.
\newblock A magic state's fidelity can be superior to the operations that
  created it.
\newblock \emph{New Journal of Physics}, 17\penalty0 (2):\penalty0 023037,
  2015.
\newblock ISSN 1367-2630.
\newblock \doi{10.1088/1367-2630/17/2/023037}.

\end{thebibliography}

%% If an appendix is needed
% \onecolumn\newpage
% \appendix
\end{document}